\documentclass[lettersize,journal]{IEEEtran}

\usepackage{amsmath,amsfonts}
\usepackage{array}
\usepackage[caption=false,font=normalsize,labelfont=sf,textfont=sf]{subfig}
\usepackage{textcomp}
\usepackage{stfloats}
\usepackage{url}
\usepackage{verbatim}
\usepackage{graphicx}
\def\BibTeX{{\rm B\kern-.05em{\sc i\kern-.025em b}\kern-.08em
    T\kern-.1667em\lower.7ex\hbox{E}\kern-.125emX}}
\usepackage{balance}

\usepackage[dvipsnames]{xcolor}
\usepackage{textgreek}
\usepackage{bbm}
\usepackage{float}
\usepackage{multirow}
\usepackage{algorithm,algcompatible}
\usepackage{amssymb}
\usepackage{amsmath}
\usepackage{amsthm}
\usepackage{graphicx}
\usepackage[numbers,sort&compress]{natbib}

\usepackage{hyperref}
\hypersetup{
    pdfborder={0 0 0}, % Turns off borders around links
}

\usepackage{glossaries}
\newacronym{3gpp}{3GPP}{3rd Generation Partnership Project}
\newacronym{bs}{BS}{base station}
\newacronym{cdf}{CDF}{cumulative distribution function}
\newacronym{ecdf}{ECDF}{empirical CDF}
\newacronym{pdf}{PDF}{probability density function}
\newacronym{ue}{UE}{user equipment}
\newacronym{los}{LoS}{line-of-sight}
\newacronym{mdt}{MDT}{minimization of drive tests}
\newacronym{ofdm}{OFDM}{orthogonal frequency division multiplexing}
\newacronym{tdoa}{TDOA}{time difference of arrival}
\newacronym{toa}{TOA}{time of arrival}
\newacronym{urllc}{URLLC}{ultra-reliable low-latency communications}
\newacronym{mar}{MAR}{maximum achievable rate}
\newacronym{pcr}{PCR}{probably correct reliability}
\newacronym{evt}{EVT}{extreme value theory}
\newacronym{gp}{GP}{Gaussian process}
\newacronym{gpd}{GPD}{generalized Pareto distribution}
\newacronym{csi}{CSI}{channel state information}
\newacronym{snr}{SNR}{signal-to-noise ratio}
\newacronym{sinr}{SINR}{signal-to-interference-plus-noise ratio}
\newacronym{mimo}{MIMO}{multiple-input multiple-output}
\newacronym{cdi}{CDI}{channel distribution information}
\newacronym{iid}{iid}{independent identically distributed}
\newacronym{mcmc}{MCMC}{Markov Chain Monte Carlo}
\newacronym{pep}{PEP}{packet error probability}
\newacronym{rssi}{RSSI}{received signal strength indicator}
\newacronym{isac}{ISAC}{integrated sensing and communication}
\newacronym{ckm}{CKM}{channel knowledge map}
\newacronym{mimi}{MIMO}{multiple-input multiple-output}
\newacronym{ai}{AI}{artifical intelligence}

\newtheorem{theorem}{Theorem}
\newtheorem{corollary}{Corollary}

\newcommand{\mbf}[1]{\mathbf{#1}}
\newcommand{\mbs}[1]{\boldsymbol{#1}}
\newcommand{\mR}{\mathbb{R}}
\newcommand*{\cond}{\hspace*{1pt} |\hspace*{1pt}}

\DeclareMathOperator{\Var}{Var}

\DeclareMathOperator*{\argmax}{arg\,max}
\newcommand{\picscale}{0.4}

\begin{document}

\title{Prediction of Rare Channel Conditions using Bayesian Statistics and Extreme Value Theory}
\author{Tobias~Kallehauge,~\IEEEmembership{Graduate Student Member,~IEEE}
        Anders~E.~Kalør,~\IEEEmembership{Member,~IEEE},
        Pablo~Ramírez-Espinosa,
        Christophe~Biscio,
        and Petar~Popovski,~\IEEEmembership{Fellow,~IEEE}% <-this % stops a space
\thanks{This work was supported in part by the Villum Investigator Grant “WATER” from the Velux Foundation, Denmark and by the industrial project ``Novel channel sounding techniques for 6G'' founded by Anritsu. The work of A.~E.~Kal{\o}r was supported by the Independent Research Fund Denmark under Grant 1056-00006B. The work by P. Ram\'irez-Espinosa has been supported by a ``Mar\'ia Zambrano" Fellowship funded by the European Union – Next Generation EU via the Ministry of Universities of the Spanish Government. }
\thanks{T. Kallehauge, A. E. Kalør and P. Popovski are with the Department of Electronic Systems, Aalborg University, Denmark (e-mails: tkal@es.aau.dk, aek@es.aau.dk, petarp@es.aau.dk).}% <-this % stops a space
\thanks{P. Ramírez-Espinosa is with the Department of Signal Theory, Networking and Communications, Universidad de Granada, Granada 18071, Spain (e-mail: pre@ugr.es)}%
\thanks{C. Biscio is with the Department of Mathematical Sciences, Aalborg University, Denmark (email: christophe@math.aau.dk)}%

}

\maketitle
\begin{abstract}
Estimating the probability of rare channel conditions is a central challenge in ultra-reliable wireless communication, where random events, such as deep fades, can cause sudden variations in the channel quality. This paper proposes a sample-efficient framework for predicting the statistics of such events by utilizing spatial dependency between channel measurements acquired from various locations. The proposed framework combines radio maps with non-parametric models and extreme value theory (EVT) to estimate rare-event channel statistics under a Bayesian formulation.
The framework can be applied to a wide range of problems in wireless communication and is exemplified by rate selection in ultra-reliable communications. Notably, besides simulated data, the proposed framework is also validated with experimental measurements. The results in both cases show that the Bayesian formulation provides significantly better results in terms of throughput compared to baselines that do not leverage measurements from surrounding locations. It is also observed that the models based on EVT are generally more accurate in predicting rare-event statistics than non-parametric models, especially when only a limited number of channel samples are available. Overall, the proposed methods can significantly reduce the number of measurements required to predict rare channel conditions and guarantee reliability.
\end{abstract}


\section{Introduction} \label{sec:intro}
Cellular network protocols depend on the availability of instantaneous \gls{csi} for beamforming algorithms, scheduling procedures, coding selection schemes, etc. However, instantaneous \gls{csi} only characterizes the channel at any moment in time and does not capture its statistical behavior, which is required to offer statistical estimates relevant to communication performance, such as expected throughput within a given time window or reliability guarantees. Specifically, systems that offer \gls{urllc} service can have very stringent latency requirements, going to below $1$ ms, such that  \gls{csi} acquisition phases may not be accommodated within the latency budget~\cite{kalor24massivecritical}. In those cases, having access to  \textit{statistical knowledge} of the channel is essential to configure the network and ensure reliability guarantees. 
However, inferring channel statistics is often challenging, particularly in ultra-reliable communication where rare events, such as deep fades, are of primary interest. For example, without model assumptions, inferring any statistics about an event that occurs with probability, e.g., $10^{-5}$, requires, on average, more than $10^5$ independent observations, which is rarely possible in practice~\cite{Angjelichinoski2019}. A more pragmatic alternative is to assume that the statistics follow some parametric distribution that can be estimated using much fewer samples. However, choosing a suitable distribution family is often challenging, and a misspecified model can lead to significant prediction errors. Thus, in practice, the distribution family often needs to be so large that estimating the parameters still requires a substantial number of samples.

A promising way to reduce the number of samples required to estimate the desired channel statistics is by leveraging the spatial correlation that is intrinsic to wireless propagation environments. This can, for instance, be done by building a \textit{radio map} to predict channel statistics at previously unobserved locations based on a database of surrounding observations in the scenario \cite{kallehauge2022Globecom,kallehauge2024experimental}. This approach has the advantage that channel statistics can be estimated fast, e.g., when a new device joins the network. 
However, spatial prediction is ultimately limited in its precision compared to methods such as in \cite{Angjelichinoski2019}, which rely on directly observing the channel at the device location. 

This paper studies how to improve existing methods for acquiring statistical channel knowledge. 
Specifically, we ask the question of \textit{how to combine radio maps with new measurements from a given location?} A natural strategy that comes to mind is to apply Bayesian inference, using the surrounding measurements as prior information, which is then updated with the new measurements to obtain a posterior estimate. However, using a small number of local measurements to aid the prediction of rare events turns out to be a non-trivial problem.
This paper proposes a novel framework that solves this problem by carefully applying Bayesian inference and by using tail-centric modeling techniques based on \textit{non-parametric} statistics and \textit{\gls{evt}}.

\subsection{Related Work and Contributions}
The problem of estimating channel statistics for rate selection in the context of \gls{urllc} was studied in \cite{Angjelichinoski2019}, using \textit{statistical learning theory} to account for the impact of finite channel observations. This work also quantifies the trade-off between non-parametric and parametric models and their respective drawbacks in the context of \gls{urllc}. The main observation is that non-parametric models require a prohibitive number of samples, while model mismatch in parametric methods can be particularly detrimental to predicting rare events. A balance for this trade-off is found in  \cite{Mehrnia2022}, which proposes to characterize small-scale fading using \gls{evt}. The authors show that, when carefully applied, \gls{evt} can significantly reduce the required number of samples compared to parametric models without penalizing the accuracy.
\Gls{evt} for \gls{urllc} has since been applied in several other works, e.g., for bandwidth and power allocation~\cite{gomes2022rare,perez2023extreme} and to construct confidence intervals for \gls{evt}-based rate selection~\cite{Mehrnia2022confidence}.

The use of spatial correlation for prediction of channel statistics has historically focused on average signal strength to construct coverage maps \cite{Chowdappa2018}. However, the increasing localization and sensing capabilities in wireless networks \cite{Lima2021Sensing} and the success of \gls{ai} have sparked a new generation of radio maps go beyond the traditional paradigm and model a plethora of channel metrics specialized for various applications~\cite{Studer2018,kallehauge2023magazine}. Examples include \textit{\gls{isac}-maps} that incorporate sensing information to predict blockages and to aid beam steering \cite{gonzalez2024integrated} and \textit{\glspl{ckm}}, a database of channel properties with spatial prediction capabilities \cite{zeng2024tutorial}. Another type of map, called \textit{\gls{cdi}-maps}, focuses specifically on channel statistics, typically small-scale fading, and has been shown as an effective tool for spatial prediction of channel statistics \cite{Kulzer2021,yu2020channel,kallehauge2022Globecom,kallehauge2024experimental,perez2024evt}. The term \gls{cdi} map was originally introduced in \cite{Kulzer2021}, which employed a dynamic framework to create a map of small-scale fading distribution parameters used for resource allocation. The work of \cite{yu2020channel} studies spatial correlation of small-scale fading under a Rician channel model. In the context of ultra-reliable communication, \cite{kallehauge2022Globecom} performs predictive rate selection for ultra-reliable communication using a \gls{cdi} map of rare-event channel statistics based on a non-parametric model, which was later on validated experimentally in \cite{kallehauge2024experimental}. 
Finally, the authors in \cite{perez2024evt} applied \gls{cdi} maps to model \gls{evt}-parameters related to rare-event statistics of the channel.

The literature review reveals two prevailing approaches for obtaining statistical channel information: \textit{(i)} directly measuring a channel in an initial training period as in \cite{Angjelichinoski2019, Mehrnia2022, Mehrnia2022confidence, gomes2022rare, perez2023extreme}, or \textit{(ii)} relying on spatial prediction through \gls{cdi} maps as in \cite{kallehauge2023magazine, kallehauge2022Globecom, perez2024evt, kallehauge2024experimental, Kulzer2021}. This work aims to unify the two approaches by proposing a framework that combines \gls{cdi} maps with new measurements in order to shed light on the previously formulated research question. Our contributions can be summarized as:
\begin{itemize}
    \item We develop a general framework for spatial statistical inference of rare events that can be used for a wide range of wireless communication scenarios. The framework is posed as a statistical learning problem with the goal of obtaining confidence intervals for rare-event quantiles of the channel that account for variability in training data.
    \item We propose two novel Bayesian estimators for the problem that enable \gls{cdi} maps to be combined with new measurements. The first relies on non-parametric statistics, while the second leverages \gls{evt}.
    \item We demonstrate how the general framework can be applied to the problem of transmission rate selection in ultra-reliable communications. This is validated numerically with both simulated and experimentally measured channels. In both cases, we show that the proposed Bayesian methods are able to estimate the achievable rate much more accurately than existing baselines that rely only on the new samples collected at the target location. 
\end{itemize}

The remainder of the paper is structured as follows. Section \ref{sec:systemmodel_problemdef} introduces the general setup of estimating rare-event channel statistics, including what information is available for inference. The proposed non-parametric approach is introduced in Sec. \ref{sec:non_para}, which also presents the framework for constructing \gls{cdi} maps. Section \ref{sec:evt} gives a short introduction to \gls{evt}, followed by the proposed \gls{evt}-based approach. The baselines are introduced in Sec. \ref{sec:baseline}, and Sec. \ref{sec:bias_discuss} discusses the impact of model bias. The scenario of rate selection is presented in Section \ref{sec:rate_selection}, where the proposed Bayesian methods are then compared to the baselines. Finally, the paper is concluded in Section \ref{sec:conclusion}.

\textit{Notation:} $\mathbb{R}$, $\mathbb{R}_{+}$, $\mathbb{C}$ denotes the sets of real, non-negative real, and complex numbers, respectively. The imaginary unit is denoted by $j$. Vectors are written in bold. $\mathcal{N}(\mu,\sigma^2)$ and $\mathcal{CN}(\mu,\sigma^2)$ denote Gaussian and circularly symmetric complex Gaussian distributions, respectively, with mean $\mu$ and variance $\sigma^2$. $E[\cdot]$ and $\Var[\cdot]$ are the expectation and variance operators, respectively. We use $\log_k$ to denote the logarithm of base $k$ and $\ln$ to denote the natural logarithm.

\textit{Reproducible research:} The code used for simulations
and figures shown in the paper can be found at: \href{https://github.com/AAU-CNT/Bayesian\_EVT\_URLLC}{https://github.com/AAU-CNT/Bayesian\_EVT\_URLLC}.

\section{Scenario and Problem Statement} \label{sec:systemmodel_problemdef}
We consider a wireless communication scenario with a single \gls{bs} at location $\mbf{s}_{\text{BS}} \in \mR^3$ that serves an area (or a cell) $\mathcal{R} \subseteq \mR^3$. A user at location $\mbf{s}_0 \in \mathcal{R}$ communicates with the \gls{bs}, and in order to guarantee reliability, it needs to characterize \textit{some} metric $X(\cdot)$ of the channel that depends on the location and its surrounding wireless propagation environment, denoted $X(\mbf{s}_0) \geq 0$.\footnote{Extensions to negative or multivariate metrics are possible but out of the scope of this paper.} Examples of $X(\mbf{s}_0)$ could include the channel \gls{snr}, the instantaneous channel coherence time, or the transmission delay, but we stress that the framework does not assume any specific metric. To study rare outcomes of the channel, this paper focuses on \textit{estimating quantiles on the tails of the channel distribution} defined as 
\begin{equation}
  X_{\epsilon}(\mbf{s}_0)=\sup\left\{x \in\mathbb{R}_+ \cond P\left(X(\mbf{s}_0) \le x \right) \le \epsilon \right\},
\end{equation}
for $\epsilon$ close to $0$ or $1$, referred to as the $\epsilon$-quantile of $X(\mbf{s}_0)$. Without loss of generality, we will focus on the lower tail of the distribution, i.e., $\epsilon$ close to $0$, which characterizes events when $X(\mbf{s}_0)$ falls significantly below its average value.

As discussed previously, a common way to estimate $X_{\epsilon}(\mathbf{s}_0)$ is to collect measurements of $X(\mathbf{s}_0)$ at $\mbf{s}_0$. However, the number of samples needed to reliably estimate $X_{\epsilon}(\mathbf{s}_0)$ can be prohibitive for small $\epsilon$, e.g., due to latency constraints or in mobile scenarios where the location of interest $\mathbf{s}_0$ changes rapidly. Instead, we aim to use spatial correlation in the propagation environment to reduce the number of samples required from the location $\mathbf{s}_0$. 
Specifically, we assume the availability of a large dataset comprising $m$ independent samples from each of $d$ locations within $\mathcal{R}$. Denoting  the set of samples from $\mathbf{s}_i$, $i=1,\ldots,d$, as $\mathbf{X}_i^m=\{X_j(\mathbf{s}_i)\}_{j=1}^m$, we denote the dataset as
\begin{equation}
  \mathcal{B}_{m,d}=\left\{\left(\mathbf{s}_i,\mathbf{X}_i^m\right)\right\}_{i=1}^d.
\end{equation}

\begin{figure}
    \centering
    \includegraphics[width = \linewidth]{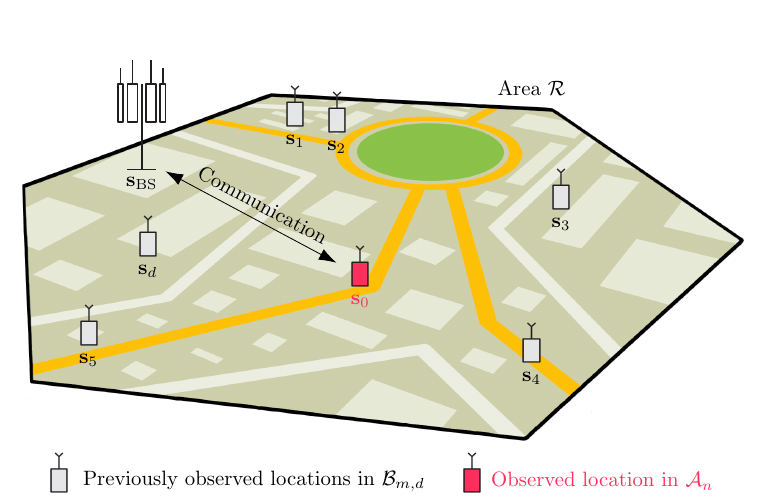}
    \caption{Communication Scenario: $d$ users in the area each previously measured the channel $m$ times giving $\mathcal{B}_{m,d}$ and the user at location $\mbf{s}_0$, whose statistics should be estimated, measured the channel $n$ times giving $\mathcal{A}_n$.}
    \label{fig:scenario}
\end{figure}

In addition to $\mathcal{B}_{m,d}$, we also assume a dataset containing $n$ of independent samples from the location of interest, $\mathbf{s}_0$, denoted as
\begin{equation}
  \mathcal{A}_{n}=\left\{\left(\mathbf{s}_0,\mathbf{X}_0^n\right)\right\}.
\end{equation}
Here, $n$ is typically small, $n < m$ or even $n \ll m$. We assume that $\mbf{s}_0$ is not among the set of locations in $\mathcal{B}_{m,d}$.
The full dataset available to estimate $X_{\epsilon}(\mbf{s}_0)$ is thus
\begin{align}
\mathcal{D} = \mathcal{A}_{n} \cup \mathcal{B}_{m,d}.\label{eq:data}
\end{align}
The scenario is illustrated in Fig. \ref{fig:scenario}. 
Note that the special case of $d = 0$ corresponds to the setting studied in \cite{Angjelichinoski2019, Mehrnia2022, Mehrnia2022confidence, gomes2022rare, perez2023extreme}, while $n = 0$ corresponds to the setting in \cite{kallehauge2023magazine, kallehauge2022Globecom, perez2024evt, kallehauge2024experimental, Kulzer2021}. Hence, having $d > 0$ and $n > 0$ unifies the two approaches. We will assume that the locations $\mbf{s}_0,\dots,\mbf{s}_d$ are perfectly known and refer the reader to \cite{kallehauge2023statistical} for a detailed discussion on the impact of localization errors in this type of scenario.

Due to the finite number of samples available, it is generally not possible to obtain a perfect estimate of $X_{\epsilon}(\mbf{s}_0)$. Inspired by the statistical learning approach in \cite{Angjelichinoski2019}, we aim instead to estimate a \emph{confidence interval} $I(\mathcal{D}) \subseteq \mR$ of $X_{\epsilon}(\mbf{s}_0)$ such that for a predefined $\delta \in (0,1)$,
\begin{align}
  P(X_{\epsilon}(\mbf{s}_0) \in I(\mathcal{D})) \geq 1 - \delta, \label{eq:general_problem_def}
\end{align}
where the probability is taken with respect to the random samples in the dataset $\mathcal{D}$. Meeting \eqref{eq:general_problem_def} provides strong guarantees on the behavior of the wireless channel by not only characterizing the quantile level for a rare event but also by accounting for limited available data.

\section{Non-Parametric Confidence Intervals} \label{sec:non_para}
We first consider non-parametric estimation of the quantile $X_{\epsilon}(\mbf{s}_0)$ following a Bayesian approach. Our proposed solution contains two steps. In the first, we use the previous measurements in the cell $\mathcal{B}_{m,d}$ to construct a probabilistic \gls{cdi} map that models the $\epsilon$-quantile across the cell. In the second step, we then use the \gls{cdi} map as a prior for the quantile at a new location $\mbf{s}_0$, and combine it with the local observations $\mathcal{A}_n$ to obtain the posterior distribution over the quantile. In the following, we first describe the general approach for obtaining probabilistic priors based on the measurements $\mathcal{B}_{m,d}$, then explain how to combine it with local observations.

\subsection{Prior Information via CDI Maps} \label{subsec:cdi_maps}
Since the prior measurements $\mathcal{B}_{m,d}$ are from different locations than the user located at $\mbf{s}_0$, they do not directly provide information about the channel of interest, so \gls{cdi} maps are used for spatial prediction. Start by denoting $\theta \in \mR$ as \textit{some variable} that we aim to predict; i.e., $\theta$ can be the quantile $X_{\epsilon}$ itself or any other parameter that may be useful. The goal is then to construct a \gls{cdi} map based on $\mathcal{B}_{m,d}$ to predict this parameter at location $\mbf{s}_0$, i.e., $\theta(\mbf{s}_0)$.

Given the observations $\mathcal{B}_{m,d}$, the first step in constructing the \gls{cdi} map is to estimate $\theta$ at the observed locations $\mbf{s}_1,\dots, \mbf{s}_d$. Letting $g$ denote a point estimator of $\theta$ based on the channel measurements, we hence compute 
\begin{align}
    \hat{\theta}(\mbf{s}_i) = g(\mbf{X}_i^m), \text{ for } i = 1,\dots, d. \label{eq:cdi_estimate_param}
\end{align}
For example, if $\theta = X_{\epsilon}$, we can use the non-parametric estimator \cite{Ord1994}
\begin{align}
    \hat{\theta}(\mbf{s}_i) = \widehat{X}_\epsilon(\mbf{s}_i) = X_{i, (r)}, \quad r = \lceil m\epsilon \rceil, \label{eq:C_out_est}
\end{align}
where $X_{i,(r)}$ is the $r$-th order statistic of $\mbf{X}_i^m$.

The next step is to construct the map via spatial prediction/interpolation, i.e., estimating $\theta(\mbf{s})$ for all $\mbf{s}$ in the cell. We use \textit{Gaussian processes} for spatial interpolation, which assumes that $\theta(\mbf{S})$ follow a multivariate Gaussian distribution for any finite subset of locations $\mbf{S} \subset \mR^3$, where the mean and covariance are determined by parametric functions of $\mbf{S}$. Gaussian processes are simple yet effective and also suit the probabilistic framework here well. Note, however, that Gaussian processes have support over all $\mathbb{R}$, and therefore non-negative variables such as $X_\epsilon$ need to be transformed. A simple way of circumventing this issue is converting to log-scale after estimation, i.e.,  $\hat{\theta}(\mbf{s}_i) = \ln(g(\mbf{X}_i^m))$ for $i = 1, \dots, d$, which shifts the range of non-negative reals into all real numbers.

For the sake of brevity, we omit the details of the prediction framework, and we refer to \cite[Sec. III-B]{kallehauge2022Globecom} for a complete description. In short, the Gaussian process framework takes the data $\{\mbf{s}_i,\hat{\theta}(\mbf{s}_i)\}_{i=1}^d$ as inputs, estimates various hyperparameters, and finally provides a \textit{predictive distribution} of $\theta$ at any location $\mbf{s}$. The predictive distribution is given by
\begin{align}
    \theta(\mbf{s}) \sim \mathcal{N}(\mu(\mbf{s}), \sigma^2(\mbf{s})), \label{eq:pred_dist}
\end{align}
where $\mu$ is the \textit{predictive mean} and $\sigma^2$ is the \textit{predictive variance}, characterizing the uncertainty of the prediction at location $\mbf{s}$. From a Bayesian perspective, the predictive distribution in \eqref{eq:pred_dist} is interpreted as the prior before collecting any samples at the target user location $\mbf{s}_0$. Note that if the logarithm transformation is applied, then $\theta(\mbf{s})$ is log-normal distributed in the linear scale. 

\subsection{Non-Parametric Bayesian Quantile Inference} \label{subsec:non_para_bay}
The non-parametric approach for Bayesian inference of the quantile is now introduced. Our strategy is to obtain the \textit{posterior distribution} of the $\epsilon$-quantile $X_{\epsilon}(\mbf{s}_0)$ from which the confidence interval can be extracted. For simplicity, we omit the dependence on the location $\mbf{s}$ from the notation in the remaining section.

The first step is obtaining a prior using a \gls{cdi} map as described in Sec. \ref{subsec:cdi_maps}. The log-scale $\epsilon$-quantile $Y_{\epsilon} \triangleq \ln(X_{\epsilon})$ is used at the statistic $\theta$ such that the prior at location $\mbf{s}_0$, denoted $f(Y_{\epsilon})$, is Gaussian with mean $\mu$ and variance $\sigma^2$. 

Next, the likelihood is inferred based on the $n$ channel measurements $\mbf{X}^n$ from location $\mbf{s}_0$ (the new samples collected at the target location). We also convert these observations to the log-scale metric, denoted $Y_{j} = \ln(X_{j})$ for $j = 1,\dots,n$. It follows that the quantile estimate  $\widehat{Y}_{\epsilon} = Y_{(r)}$ with $r = \lceil n\epsilon\rceil$ is unbiased and asymptotically normally distributed as \cite{Ord1994}
\begin{align}
    \widehat{Y}_{\epsilon} - Y_{\epsilon} \overset{d}{\to} \mathcal{N}\left(0, \frac{\epsilon(1-\epsilon)}{nf_Y(Y_{\epsilon})^2} \right), \label{eq:normal_quantile}
\end{align}
where $\overset{d}{\to}$ denotes convergence in distribution and $f_Y$ is the \gls{pdf} of $Y = \ln(X)$. In the Bayesian framework, we say that $\widehat{Y}_{\epsilon} | Y_{\epsilon}$ is Gaussian distributed with mean $Y_{\epsilon}$ and variance $\sigma_n^2 = \epsilon(1-\epsilon)/(nf_Y(Y_{\epsilon})^2)$ with likelihood denoted by $f(\widehat{Y}_{\epsilon} | Y_{\epsilon})$. Assuming for now that the variance $\sigma_n^2$ is known, we get the posterior distribution from Bayes theorem
\begin{align}
    f(Y_{\epsilon} \cond \widehat{Y}_{\epsilon}) = \frac{f(\widehat{Y}_{\epsilon} | Y_{\epsilon})f(Y_{\epsilon})}{f(\widehat{Y}_{\epsilon})},
\end{align}
where the numerator is the \textit{evidence} given by $f(\widehat{Y}_{\epsilon}) = \int f(\widehat{Y}_{\epsilon} | Y_{\epsilon})f(Y_{\epsilon}) \ d Y_{\epsilon}$. Note that both the prior and likelihood are Gaussian, and thus the posterior $f(Y_{\epsilon} \cond \hat{Y}_{\epsilon})$ is also Gaussian with \cite{Kay1998}
\begin{align}
    \mu_{\text{post}} &= E[Y_{\epsilon} \cond \widehat{Y}_{\epsilon}] = \frac{\sigma_n^2}{\sigma_n^2 + \sigma^2}\mu + \frac{\sigma^2}{\sigma_n^2 + \sigma^2}\widehat{Y}_{\epsilon}, \label{eq:post_mean} \\
    \sigma^2_{\text{post}} &= \Var[Y_{\epsilon} \cond \hat{Y}_{\epsilon}] = \left(\frac{1}{\sigma_n^2} + \frac{1}{\sigma^2} \right)^{-1}. \label{eq:post_var}
\end{align}
Since the posterior distribution of $Y_{\epsilon}$ is Gaussian, the posterior for $X_{\epsilon}$ is log-normal whose inverse \gls{cdf} is given by
\begin{align}
    F_{X_\epsilon}^{-1}(p) = \exp\left(\mu_{\text{post}} + \sqrt{2}\sigma_{\text{post}}\text{erf}^{-1}(2p - 1)\right),
\end{align}
for $p \in [0,1]$, where $\mathrm{erf}^{-1}$ is the inverse error function. It follows that a confidence interval with confidence level $1 - \delta$ can be obtained either as a one or two-sided interval:
\begin{align}
    I(\mathcal{D}) &= [F_{X_\epsilon}^{-1}(\delta), \infty) \quad (\text{one-sided}) \label{eq:non_para_bay_one_sided}, \\
    I(\mathcal{D}) &= [F_{X_\epsilon}^{-1}(\delta/2), F_{X_\epsilon}^{-1}(1 - \delta/2)] \quad (\text{two-sided}). \label{eq:non_para_bay_two_sided}
\end{align}
Inserting either \eqref{eq:non_para_bay_one_sided} or \eqref{eq:non_para_bay_two_sided} into \eqref{eq:general_problem_def} gives a probability of exactly $1 - \delta$ of containing the true quantile $X_{\epsilon}$ assuming that the prior is correct, that the asymptotic result in \eqref{eq:normal_quantile} holds, and that $\sigma_n^2$ is known. These assumptions will be violated to some degree in practice, and $\sigma_n^2$ needs to be estimated as well. We found that good performance can achieved by spatial prediction of $\sigma_n^2$ using a separate \gls{cdi} map as described in Sec. \ref{subsec:cdi_maps} with $\theta = f_Y(Y_{\epsilon})$, and then using the point estimate ${\hat{\sigma}_n^2 = \epsilon(1-\epsilon)/(n\hat{f}_Y(Y_{\epsilon})^2)}$. Note that it is also possible to directly incorporate uncertainty about $\sigma_n^2$ in the Bayesian framework, but at the cost of losing analytic tractability, so we choose to use the point estimate and ignore its uncertainty. 

\section{EVT-Based Confidence Intervals}
In this section, we present a Bayesian approach for quantile inference using \gls{evt} as an alternative to the non-parametric model. As in the previous section, we rely on a \gls{cdi} map to model the prior of the \gls{evt} parameters and then use the local observations to obtain a posterior distribution of the quantile. We start by introducing the relevant theory and then explain how to apply it for quantile inference. 

\subsection{EVT for Threshold Deficits} \label{subsec:evt_deficit}
\Gls{evt} is the study of extreme outcomes of random events, providing parametric distributions that apply to \textit{any} random event and can be applied to a large range of distributions \cite{coles2001introduction}. The following theorem, which is one of the core results in \gls{evt}, is useful for the problem at hand, as it provides a conditional distribution for values below a threshold through the \gls{gpd}.
\begin{theorem}[Pickands–Balkema–De Haan theorem for deficits~\cite{balkema1974residual, Mehrnia2022}] \label{thm:pickands}
Let $X$ be a random variable following some distribution $F$ and let $F_u$ be the conditional distribution function for the deficit $Y = u - X$ given $X < u$ for some threshold $u$, i.e., 
\begin{align}
    F_u(y) = P(u - X \leq y \cond X < u).
\end{align}
Then, for sufficiently low $u$ and for a wide range of distributions $F$, $F_u$ is approximately a \gls{gpd} with \gls{cdf}
\begin{align}
    F_u(y; \sigma_u, \xi) = 1 - \left(1 + \frac{\xi y}{\sigma_u} \right)^{-1/\xi},
\end{align}
and \gls{pdf}
\begin{align}
    f_u(y; \sigma_u, \xi) = \frac{1}{\sigma_u}\left(1 + \frac{\xi y}{\sigma_u} \right)^{-(1 + 1/\xi)} \label{eq:gpd_pdf}
\end{align}
defined on the domain $\{y > 0 \wedge 1 +  \xi y/\sigma_u > 0\}$, where $\sigma_u > 0$ is the scale parameter and $\xi \in \mR$ is the shape parameter.
\end{theorem}
Theorem \ref{thm:pickands} notably applies under very mild conditions on $X$ with the conditional distribution of the deficit $Y$ converging to the \gls{gpd} as $u \to -\infty$. We refer the reader to \cite{coles2001introduction} for the specific conditions and further details on \gls{evt} and \cite{balkema1974residual} for the original proof by Balkema and De Haan. Since we are interested in quantiles on the lower tail of the channel distribution, it is likely that the \gls{gpd} can provide a good approximation that allows us to characterize the $\epsilon$-quantile $X_{\epsilon}$ without imposing strong assumptions on the distribution of $X$.

A parametric form for the tail distribution and quantiles follows from Thm.~\ref{thm:pickands}. 
\begin{corollary}[Tail distribution and quantiles~\cite{coles2001introduction, Mehrnia2022}]
    Denote by $F_{\text{tail}}$ the tail distribution of $X$, i.e., $F_{\text{tail}}(x) = P(X \leq x)$ for $x<u$. Given \gls{gpd} parameters for the deficit $(\sigma_u, \xi, u)$ it follows that
    \begin{align}
    F_{\text{tail}}(x; \sigma_u, \xi, u) = \left(1 + \frac{\xi(u - x)}{\sigma_u} \right)^{-1/\xi}\cdot p_u, \label{eq:gpd_tail}
\end{align}
where $p_u = P(X \leq u)$. The quantile $X_\epsilon = F_X^{-1}(\epsilon)$ follows directly from \eqref{eq:gpd_tail} as
\begin{align} 
    X_{\epsilon} = u - \frac{\sigma_u}{\xi}\left(\left(\frac{p_u}{\epsilon}\right)^{\xi} - 1\right), \label{eq:GPD_quantile}
\end{align}
provided that $X_{\epsilon} \leq u$, with $F_X$ the \gls{cdf} of $X$ and $\epsilon\in(0,1]$.
\end{corollary}

Given $u$, the parameters of the \gls{gpd} can be estimated using conventional estimation techniques. Given independent observations $\mbf{X}^n = \{X_i\}_{i=1}^n$, the deficit is computed as $\mbf{Y} = {\{u - X_i \cond X_i < u\}}$. The parameters can then be estimated from $\mbf{Y}$, e.g., through maximum likelihood estimation as
\begin{align}
    (\hat{\sigma}_u,\hat{\xi}) = \argmax_{\sigma_u,\xi} \prod_{y \in \mbf{Y}} f_u(y; \sigma_u, \xi), \label{eq:mle_gpd}
\end{align}
where $f_u$ is the \gls{pdf} given in \eqref{eq:gpd_pdf}.

\subsection{Threshold Selection} \label{subsec:threshold}
The threshold $u$ determines from which point the \gls{gpd} is a good model for the deficit of the random process $X$, and we have the result that if Thm. \ref{thm:pickands} applies for some threshold $u_0$, it also applies for all $u \leq u_0$ \cite{coles2001introduction}. The Bayesian \gls{evt}-based approach introduced in the following requires the threshold $u$ to be selected for each of the observed locations $\mbf{s}_0,\dots, \mbf{s}_d$ based on the measurements in the dataset $\mathcal{D}$. A popular method for threshold selection is the \textit{mean deficit plot}, which is simple and generally accurate but typically requires a manual decision of the threshold \cite{scarrott2012review}. Another method is \textit{fixed threshold}, where the threshold is based on order statistics. Given observations $\mbf{X}_i^m$ at location $\mbf{s}_i$, the fixed threshold approach selects $u_i = X_{i,(r)}$ for some $r \in [1,m]$ \cite{scarrott2012review}. A simple rule that has been applied in the literature is to pick $r = m/10$ such that $10\%$ of the observed data is used to model the deficit \cite{scarrott2012review}. Choosing the threshold based on order statistics allows for direct control of the bias/variance tradeoff of fitting the \gls{gpd}. When $r$ is low, the \gls{gpd} is more likely to be a good model of the deficit (low bias), but there are fewer samples to fit the distribution (high variance) and vice-versa for a high $r$. The fixed threshold is also easier to automate than the mean deficit plot method but can be less accurate if the choice of $r$ is not suitable for the actual data. Based on these considerations, we adopt the fixed threshold selection technique in this paper with certain modifications to improve accuracy. Fixed threshold also have certain analytic properties that we will exploit later. 

Given $m$ observations of the channel $\mbf{X}_{i}^m$ at location $\mbf{s}_i$ (or $n$ observations at $\mbf{s}_0$), the modified threshold selection rule is given by
\begin{align}
    u_i = X_{i,(r)}, \quad r = \max(\lceil n\zeta \rceil, r_{\min}), \label{eq:threshold_zeta}
\end{align}
where $\zeta \in (\epsilon, 1]$ denotes the fraction of samples used to model the tail distribution when $\lceil n \zeta \rceil > r_{\min}$ and $r_{\min}$ is the minimum number of samples used to model the tail. This approach enables automatic threshold selection given the hyperparameters $\zeta$ and $r_{\min}$.
%We found that setting $r_{\min} = 200$ greatly reduces the variance of fitting the \gls{gpd} when $n$ is low. 
Regarding $\zeta$, we observed that it tends to be in a similar range for channels in similar environments. We also found that the specific choice of $\zeta$ does not have a significant impact on the accuracy of the \gls{gpd} for the deficit, e.g., changing $\zeta$ relatively within $\pm 50\%$ has only a minor effect on the inference framework presented below. Thus, a global $\zeta$ can then be selected for all channels within the cell. Appendix \ref{subsec:zeta_heuristic} introduces a heuristic approach to choosing a global $\zeta$ based on the mean deficit plot. 

\subsection{EVT-Based Bayesian Quantile Inference} \label{subsec:evt_bayesian} 
Given the threshold selection rule in \eqref{eq:threshold_zeta}, inference of the $\epsilon$-quantile is divided into three steps: First, obtaining a prior of \gls{gpd} parameters via \gls{cdi} maps; secondly, inferring the posterior distribution using \textit{\gls{mcmc}} methods; and finally, obtaining the confidence interval based on the posterior. To avoid the dependence on the threshold $u$ of $\sigma_u$, which would require the same value of $u$ at all locations, we reparametrize from $(\sigma_u, \xi, p_u)$ to $(X_{\epsilon}, \xi, p_u)$ which have a one-to-one relation following the result in \eqref{eq:GPD_quantile}. This parametrization is also useful as it directly contains the $\epsilon$-quantile $X_{\epsilon}$, which is the parameter of interest. Of course, $p_u$ also clearly depends on the threshold, but we shall see that its prior can be obtained without utilizing spatial prediction. 

With the prior measurements $\mathcal{B}_{m,d}$ and user location $\mbf{s}_0$, the first step is then to obtain a prior for the parameters denoted jointly as $\mbs{\phi} = (X_{\epsilon}, \xi, p_u)$ (again, the location $\mbf{s}_0$ is omitted for notational simplicity). A prior for the $\epsilon$-quantile is obtained as in Sec. \ref{subsec:cdi_maps} using a \gls{cdi} map with $\theta = Y_{\epsilon} = \ln(X_{\epsilon})$ such that the prior for $X_{\epsilon}$ at user location $\mbf{s}_0$ is log-normal (in the liner domain), denoted as $f(X_{\epsilon})$. Similarly, a prior for $\xi$ is also obtained using \gls{cdi} maps with $\theta = \xi$, using now \eqref{eq:mle_gpd} as the estimator in \eqref{eq:cdi_estimate_param}. 
In this case, the log-transformation trick is not necessary since $\xi$ can take any real value, including negatives\footnote{In theory, the \gls{gpd} model for the deficit of the channel should always have $\xi < 0$ since otherwise it would imply a non-zero probability for observing negative $X$, which was assumed non-negative. However, in practice, the \gls{gpd} is only applied in a small range of values, and we found that letting the parameter vary freely yields the best results.}. The prior for $\xi$ is thus Gaussian, denoted as $f(\xi)$. The prior for $p_u$ is obtained by exploiting the threshold rule in \eqref{eq:threshold_zeta}. Since $u_i = X_{i,(r)}$ for each $i = 1,\dots, d$, it follows that $p_u = P(X_i \leq X_{i,(r)})$ is beta-distributed with shape parameters $r$ and $n + 1 - r$ \cite{Ord1994}, denoted $f(p_u)$. The joint parameter prior is then $f(\mbs{\phi}) = f(X_{\epsilon})f(\xi)f(p_u)$, i.e., the product of a log-normal, a normal, and a beta distribution. 

The next step is inferring the posterior distribution for the $\epsilon$-quantile by combining a likelihood with the prior. For the likelihood, we use the $n$ observations $\mbf{X}_0^n$ at the user location $\mbf{s}_0$. The threshold $u$ is firstly selected as in \eqref{eq:threshold_zeta} and the deficits are then computed as  $\textbf{Y} = \{y_i\}_{i=1}^{r-1} = \{u - X_{0,i} \cond X_{0,i} < u\}$, where the number of observations are $|\mbf{Y}| = r - 1 =  \lceil n\zeta\rceil - 1$. The likelihood for $\mbs{\phi} = (X_{\epsilon},\xi,p_u)$ is then
\begin{align}
   f(\mbf{Y} \cond X_{\epsilon},\xi,p_u) &= \prod_{i=1}^{r-1} f_u(y_i; \sigma_u(X_{\epsilon}, \xi, p_u), \xi) \label{eq:like_gpd}\\
    \sigma_u(X_{\epsilon}, \xi, p_u) &= \frac{(u - X_{\epsilon})\xi}{\left(\frac{p_u}{\epsilon}\right)^{\xi} -1}, \label{eq:reparametrization}
\end{align}
where $f_u$ is the \gls{pdf} given in \eqref{eq:gpd_pdf} and \eqref{eq:reparametrization} follows from \eqref{eq:GPD_quantile}. Bayes theorem now gives that 
\begin{align}
    f(X_{\epsilon},\xi,p_u \cond \mbf{Y}) \propto f(\mbf{Y} \cond X_{\epsilon},\xi,p_u) f(X_{\epsilon})f(\xi)f(p_u), \label{eq:post_GPD}
\end{align}
i.e., the posterior is proportional to the likelihood multiplied by the prior. Unlike in the non-parametric example, where the posterior was analytically tractable, the posterior in \eqref{eq:post_GPD} requires numerical approximation. We here use the \gls{mcmc} method known as \textit{metropolis within Gibbs} that enables drawing random samples from $\mbs{\phi}  \cond \mbf{Y}$ only based on the unscaled posterior (right-hand side of \eqref{eq:post_GPD}), thus avoiding having to compute the proportionality constant. The method is summarized in Algorithm \ref{alg:metropolis_within_gibbs} wherein the parameters are updated sequentially by proposing a new value based on the previous (line 2, 9 and 16) and then accepting the new proposal with probability $\min(A,1)$ where $A$ is the \textit{metropolis ratio} (line 3, 10 and 17) --- we refer the reader to the \href{https://github.com/AAU-CNT/Bayesian\_EVT\_URLLC}{supplementary resources} for additional implementation details and to \cite[ch. 11]{bishop2006pattern} for background on \gls{mcmc} methods. 
\begin{algorithm}[t]
\caption{Metropolis within Gibbs}\label{alg:metropolis_within_gibbs}
\begin{algorithmic}[1]\small
\REQUIRE Threshold $u$, deficits $\mbf{Y}$, parameters for the prior of $\mbs{\phi}$, initial parameter $\mbs{\phi}_0 = (X_{\epsilon,0}, \xi_0, p_{u,0})$, proposal variances $\mbf{s} = (s_{X_\epsilon},s_{\xi},s_{p_u})$, and number of iterations $T$.
\FORALL{$t = 1,\dots, T$}
%     \STATEx \quad \textit{First update $X_{\epsilon}$}
     \STATE $\widetilde{X}_{\epsilon} \sim \mathcal{N}(X_{\epsilon,t-1}, s_{X_\epsilon})$, \quad $U\sim \text{Unif}(0,1)$
     \STATE $A \gets \dfrac{f(\mbf{Y} \cond \widetilde{X}_{\epsilon},\xi_{t-1},p_{u,t-1}) f(\widetilde{X}_{\epsilon})}{f(\mbf{Y} \cond X_{\epsilon,t-1},\xi_{t-1},p_{u,t-1}) f(X_{\epsilon,t-1})}$
     \IF{$U \leq \min(A,1)$}
        \STATE $X_{\epsilon,t} \gets \widetilde{X}_{\epsilon}$
     \ELSE
        \STATE $X_{\epsilon,t} \gets X_{\epsilon,t-1}$
     \ENDIF
     
     \STATEx \quad \textit{Then update $\xi$}
     \STATE $\tilde{\xi} \sim \mathcal{N}(\xi_{t-1}, s_{\xi})$, \quad $U\sim \text{Unif}(0,1)$
     \STATE $A \gets \dfrac{f(\mbf{Y} \cond X_{\epsilon,t},\tilde{\xi},p_{u,t-1}) f(\tilde{\xi})}{f(\mbf{Y} \cond X_{\epsilon,t},\xi_{t-1},p_{u,t-1}) f(\xi_{t-1})}$
     \IF{$U \leq \min(A,1)$} 
        \STATE $\xi_t \gets \tilde{\xi}$
     \ELSE
        \STATE $\xi_t \gets \xi_{t-1}$
     \ENDIF

     \STATEx \quad \textit{Finally update $p_u$}
     \STATE $\tilde{p}_u \sim \mathcal{N}(p_{u,t-1}, s_{\xi})$, \quad $U\sim \text{Unif}(0,1)$
     \STATE $A \gets \dfrac{f(\mbf{Y} \cond X_{\epsilon,t},\xi_t,\tilde{p}_u) f(\tilde{p}_u)}{f(\mbf{Y} \cond X_{\epsilon,t},\xi_{t},p_{u,t-1}) f(p_{u,t-1})}$
     \IF{$U \leq \min(A,1)$} 
        \STATE $p_{u,t} \gets \tilde{p}_u$
     \ELSE
        \STATE $p_{u,t} \gets p_{u,t-1}$
     \ENDIF    
\ENDFOR
\end{algorithmic}
\end{algorithm}
The output of Alg. \ref{alg:metropolis_within_gibbs} are $T$ simulated parameters $\mbs{\phi}_1,\dots,\mbs{\phi}_T$ from the posterior distribution and among these $T$ values of the $\epsilon$-quantile $X_{\epsilon,1},\dots,X_{\epsilon,T}$. 

Finally, the confidence interval for the $\epsilon$-quantile can be inferred from the simulated Markov chain. The inverse \gls{cdf} of the posterior distribution of $X_{\epsilon}$ is approximated by the order statistic
\begin{align}
    F_{X_{\epsilon}}^{-1}(p) \approx X_{\epsilon, (\lceil pT \rceil)}, \label{eq:chain_ICDF}
\end{align}
for $p \in [0,1]$, where the approximation can be made arbitrarily tight by increasing the number of iterations $T$. %We use $T = 10^4$ for the simulations in Sec. \ref{sec:rate_selection}. 
The confidence interval $I(\mathcal{D})$ is then found by using \eqref{eq:chain_ICDF} in \eqref{eq:non_para_bay_one_sided} for a one-sided interval or \eqref{eq:non_para_bay_two_sided} for a two-sided interval.

 \label{sec:evt}

\section{Baselines} \label{sec:baseline}
To illustrate the advantages of incorporating prior information with the Bayesian frameworks, two baselines that solely rely on the observations at the new user location (i.e., $d = 0$) are also introduced. The available data for the baselines are then $\mathcal{A}_{n}$, i.e., the $n$ channel measurements $\mbf{X}^n = \{X_i\}_{i=1}^n$ taken at the target location $\mbf{s}_0$ --- we drop again the dependence on $\mbf{s}_0$ in the following. One baseline will use a non-parametric model, and one will use \gls{evt}.
\subsection{Non-Parametric Baseline}
This baseline used a non-parametric inference and was first introduced in \cite{Angjelichinoski2019}. Let $X_{(r)}$ denote the $r$-th order statistic of $\mbf{X}^n$ for $r \in \{1,\dots,n\}$. Then \cite{Ord1994}
\begin{align}
    P(X \leq X_{(r)}) = F_X(X_{(r)}) \sim \text{Beta}(r,n+1-r) \label{eq:beta_orderstat}
\end{align}
where $F_X$ is the \gls{cdf} of the channel $X$ such that
\begin{align}
    P( F_X(X_{(r)}) \leq p) = I_{p}(r, n+ 1 - r)
\end{align}
for $p \in [0,1]$, where $I_{p}$ is the \gls{cdf} of the beta distribution given by the regularized incomplete beta function. Noting also that the $\epsilon$-quantile is given by the inverse \gls{cdf} $X_{\epsilon} = F_X^{-1}(\epsilon)$, it follows that a one-sided confidence interval can be constructed as $I(\mathcal{A}_n) = [X_{(r)}, \infty)$ by observing that
\begin{align}
    P(X_{\epsilon} \in I(\mathcal{A}_n)) &= P(X_{\epsilon} \leq \infty) - P(X_{\epsilon} \leq X_{(r)}) \nonumber \\
    &= 1 - P(F_X^{-1}(\epsilon) \leq X_{(r)}) \nonumber \\
    &= 1 - P(\epsilon \leq F_X(X_{(r)})) \nonumber \\
    &= P(F_X(X_{(r)}) < \epsilon) \nonumber \\
    &= I_{\epsilon}(r,n+1-r). \label{eq:non_para_analytical}
\end{align}
Selecting $r$ as the maximum value such that $I_{\epsilon}(r,n+1-r) \geq 1 - \delta$ then ensures that \eqref{eq:general_problem_def} is fulfilled. A two-sided interval can be constructed following a similar approach. Given $I(\mathcal{A}_n) = [X_{(r_1)}, X_{(r_2)}]$, it follows that 
\begin{align}
    P(X_{\epsilon} \in I(\mathcal{A}_n)) = I_{\epsilon}(r_1,n+1-r_1) - I_{\epsilon}(r_2,n+1-r_2). \label{eq:two_sided_interval_baseline_nonpara}
\end{align}
Setting $r_1 = \lceil n\epsilon \rceil - k$ and $r_2 = \lceil n\epsilon \rceil + k$, the confidence interval fulfills \eqref{eq:general_problem_def} when $k$ is selected as the minimum value such that the right-hand-side of \eqref{eq:two_sided_interval_baseline_nonpara} exceeds $1-\delta$. Remarkably, this method does not depend on the actual channel distribution, and unlike the previous methods, the statistical guarantee is exact since no asymptotic results were used. However, we shall see in the results that this comes at the cost of confidence intervals that are generally larger than Bayesian or parametric approaches due to the lack of prior knowledge about the channel. 

\subsection{EVT Baseline} \label{subsec:evt_baseline}
This baseline combines \gls{evt} with \textit{profile likelihood} inference inspired by \cite[p. 81-83]{coles2001introduction}. Similar to Sec. \ref{subsec:evt_bayesian}, the first step is to compute the threshold $u$ according to \eqref{eq:threshold_zeta} given the channel measurement $\mbf{X}^n$, and then obtaining the deficits $\textbf{Y} = \{y_i\}_{i=1}^{r-1}$. We again use the \gls{gpd} parametrization in terms of $(X_{\epsilon}, \xi, p_u)$. However, we directly plug-in the estimate $\hat{p}_u = r/n$ suggested in \cite{coles2001introduction}, 
which greatly reduces the computational complexity for the inference strategy introduced below. The \textit{log-likelihood} of the deficit is then 
\begin{align}
    \ell(X_{\epsilon}, \xi; \mbf{Y}) = \sum_{i=1}^{r-1} \ln(f_u(y_i; \sigma_u(X_{\epsilon}, \xi, \hat{p}_u), \xi), \label{eq:log_like_GPD}
\end{align}
where $f_u$ is the \gls{pdf} in \eqref{eq:gpd_pdf} and $\sigma_u$ is given in \eqref{eq:reparametrization}. The log-likelihood can be used to infer $X_{\epsilon}$, but the presence of the nuisance parameter $\xi$ may cause bad small-sample properties of the confidence interval. The idea of the profile likelihood is then to eliminate nuisance parameters, replacing them with their maximum-likelihood estimates. In this case, the (log) profile likelihood is obtained by
\begin{align}
    \ell_p(X_{\epsilon}; \mbf{Y}) = \sup_{\xi}  \ell(X_{\epsilon}, \xi; \mbf{Y}). 
\end{align}
While the profile likelihood is not a true likelihood function, we can construct the confidence interval
\begin{equation}
    \resizebox{\linewidth}{!}{
        $ I(\mathcal{A}_n) = \left\{X_\epsilon \geq 0: 2\left(\ell(\widehat{X}_{\epsilon}, \hat{\xi}; \mbf{Y}) - \ell_p(X_{\epsilon}; \mbf{Y})\right) \leq \mathcal{X}_{1 - \delta}^2(1)\right\}$, \label{eq:conf_gdp_freq}
    }
\end{equation}
where $(\widehat{X}_{\epsilon}, \hat{\xi})$ are the maximum likelihood estimates of $(X_{\epsilon},\xi)$ obtained using \eqref{eq:log_like_GPD} and $\mathcal{X}_{1-\delta}^2(1)$ is the $1-\delta$ quantile of the chi-squared distribution with one degree of freedom. The confidence interval in \eqref{eq:conf_gdp_freq} is asymptotically tight, i.e., $P(X_{\epsilon} \in I(\mathcal{A}_n)) \overset{d}{\to} 1-\delta$, and generally has better small-sample accuracy than likelihood-based intervals  \cite{coles2001introduction}. The interval obtained using the profile likelihood tends to be two-sided, denoted $I(\mathcal{A}_n) = [I_{\min}(\mathcal{A}_n), I_{\max}(\mathcal{A}_n)]$. A one-sided confidence interval can be constructed from the two-sided one simply as $I(\mathcal{A}_n) = [I_{\min}(\mathcal{A}_n), \infty)$, which still fulfills \eqref{eq:general_problem_def}, but at the cost of producing a slightly larger interval than required.

\section{Limiting Behavior of Coverage Guarantees in the Presence of Model Bias} \label{sec:bias_discuss}
We conclude the theoretical part by discussing the behavior of the coverage guarantee $P(X_{\epsilon} \in I(\mathcal{D}))$, which is ideally close to $1 - \delta$, as the number of observations $n$ at the new user location $\mathbf{s}_0$ increase. In particular, we analyze the impact of \textit{model bias}. 

Let us first assume a one-sided confidence interval $I_n(\mathcal{D}) = [I_{\min,n}
(\mathcal{D}), \infty)$, where we have explicitly included $n$ in the notation to emphasize the dependency on $n$. As $n$ increases, it is reasonable to assume that the confidence interval converges. Formally, we assume that $I_{\min,n}(\mathcal{D})$ converges in probability, denoted $I_{\min,n}(\mathcal{D}) \overset{p}{\to} I_{\min}$ for $n \to \infty$ where $I_{\min} > 0$ is the limit for the lower bound. We can always write $I_{\min} = X_{\epsilon} + b$ where $b$ is the \textit{bias} with respect to the true $\epsilon$-quantile $X_{\epsilon}$. A bias can occur if the model used to infer the confidence interval differs from the true probability distribution of $X$,
%and we say that the confidence interval is \textit{inconsistent}
which will be the case in most practical scenarios. In particular, while the \gls{gpd} models used in Sec. \ref{subsec:evt_bayesian} and \ref{subsec:evt_baseline} are asymptotically correct, there will be some bias for any constant choice of $\epsilon$. Now since, $X_{\epsilon} \in I_n(\mathcal{D}) \Leftrightarrow {I_{\min,n}(\mathcal{D}) - X_{\epsilon} < 0}$ for a one-sided interval, and since  $I_{\min,n} \overset{p}{\to} I_{\min}$ implies ${I_{\min,n}(\mathcal{D}) - X_{\epsilon}}\overset{p}{\to} I_{\min} - X_{\epsilon} = b$, it follows directly that 
\begin{align}
        b > 0 &\Rightarrow \lim_{n\to \infty} P(X_{\epsilon} \in I_n(\mathcal{D})) = 0 \label{eq:statement_positive_bias} \\
        b < 0 &\Rightarrow \lim_{n\to \infty} P(X_{\epsilon} \in I_n(\mathcal{D})) = 1 \label{eq:statement_negative_bias},
    \end{align}
that is, the coverage guarantee converges to either $0$ or $1$ depending on a positive or negative bias, respectively. Intuitively, this is because, as the lower bound $I_{\min,n}(\mathcal{D})$ converges to the limit $I_{\min}$, any non-zero bias will cause the lower bound to either under or overestimate the quantile $X_{\epsilon}$. Similarly, for a two-sided confidence interval $I_{n}(\mathcal{D}) = [I_{\min,n}(\mathcal{D}),I_{\max,n}(\mathcal{D})]$, it follows that if $I_{n}(\mathcal{D})$ converges in probability to a point $I \in \mR$, then any non-zero bias $b = I - X_{\epsilon} \neq 0$ will cause $P(X_{\epsilon} \in I_n(\mathcal{D})) \to 0$ for $n \to \infty$.

There are some key takeaways from this discussion. First, providing a correct coverage guarantee is generally difficult in the high-sample regime since any model bias will tend to dominate as the number of observations grows. Second, non-parametric methods can avoid the bias problem by not relying on a model, which is an argument for using non-parametric models in the high-sample domain.  Finally, although even a small bias $b \approx 0$ may cause the coverage guarantee to be violated, the estimated confidence interval can still be accurate. For example, in the one-sided case, the estimation error fulfills $|X_{\epsilon} - I_{\min,n}| \overset{p}{\to} b \approx 0$ given the previous assumptions. We will illustrate these observations numerically in Sec. \ref{sec:rate_selection}.

\section{Location-Aided Rate Selection for Ultra-Reliable Communication} \label{sec:rate_selection}
The general framework of inferring confidence intervals for the $\epsilon$-quantile of the channel is now used for the problem of selecting a rate in ultra-reliable communications that accounts for rare events of the wireless channel (i.e., fading process). We start by introducing the signal model and rate selection problem and then evaluate the proposed Bayesian methods compared to the baselines --- first through simulated and then through experimentally measured channels. 

\subsection{Setup and Signal Model}

The setup considered here is a specific case of the general scenario in Section \ref{sec:systemmodel_problemdef}, where a new user at location $\mbf{s}_0$ within the cell initializes an uplink transmission, e.g., an existing user moving to a new location or a user joining the network for the first time. Prior to the transmission, the user is able to acquire $n$ channel measurements at $\mbf{s}_0$ to assist the estimation of channel statistics. Note that these samples do not need to be collected immediately before the transmission but take advantage of, e.g., idle moments. 

A narrowband transmission is assumed, and both \gls{ue} and \gls{bs} are equipped with a single antenna each.
Interference from within and outside the cell is also neglected here. We stress that these assumptions, while easing the presentation, are not required for the rate-selection approach, which is independent of the specific channel model. Following this, normalized uplink signals $\mathbf{a}\in\mathbb{C}^l$ such that $E[\|\mathbf{a}\|_2^2]=l$ are transmitted by the \gls{ue} and received as
\begin{equation}
    \mathbf{y} = h\mathbf{a} + \mathbf{z}, \label{eq:rx_signal}
\end{equation}
where $h$ is the complex channel gain and $\mathbf{z}\in\mathbb{C}^l$ is the noise vector with elements drawn independently from $\mathcal{CN}(0,BN_0)$ with $N_0$ the power spectral density and $B$ the transmission bandwidth. The distribution of $h$ is imposed by the propagation environment, which is assumed stationary and is thus constant but unknown for a given location $\mbf{s}$. The instantaneous \gls{snr} at the \gls{bs} is therefore
\begin{equation}
    \gamma =\frac{|h|^2}{BN_0}. \label{eq:snr}
\end{equation}
At the physical layer, the reliability is given by the \gls{pep}, which is well approximated by the \textit{outage probability}, even for moderately short blocklengths $l$ \cite{yang14quasistatic}, and is given by
\begin{equation}
    p_{\text{out}}(R) = P(\log_2(1+\gamma) < R) = F_{C}(R), \label{eq:Pout}
\end{equation}
where $R$ is the transmission rate, and $F_{C}$ is the \gls{cdf} of the instantaneous capacity $C = \log_2(1+\gamma)$. Since the dynamics of $C$ are directly linked to reliability, this will be used as the aforementioned channel metric of interest, i.e., $X = C$. The \textit{$\epsilon$}-outage capacity is the largest rate that the channel can support with probability $1 - \epsilon$, i.e., 
\begin{align}
    C_{\epsilon} = \sup_{R}\{R \geq 0 \cond P(\log_2(1+\gamma) \leq R) < \epsilon\}, \label{eq:C_out}
\end{align}
and is exactly the $\epsilon$-quantile of $C$.

\subsection{Rate Selection Problem} \label{sec:rate_selection_problem}
We pose the problem of selecting the communication rate $R$ given the available data 
$\mathcal{D}$ in \eqref{eq:data} (with $X = C$) for the user at the location $\mathbf{s}_0$ while ensuring that the outage probability in \eqref{eq:Pout} stays below some target $\epsilon \in (0,1)$. It follows from \eqref{eq:C_out} and \eqref{eq:Pout} that the outage probability fulfills $p_{\text{out}}(R) \leq \epsilon$ if and only if the rate fulfills $R \leq C_{\epsilon}$. Hence, given a one-sided confidence interval $I(\mathcal{D}) = [I_{\min}(\mathcal{D}), \infty)$ for the outage capacity $C_{\epsilon}$ that fulfills \eqref{eq:general_problem_def}, and by selecting $R(\mathcal{D}) = I_{\min}(\mathcal{D})$, we see that
\begin{align}
    P(p_{\text{out}}(R(\mathcal{D})) \leq \epsilon) &= P(C_{\epsilon} \geq R(\mathcal{D})) \nonumber  \\
    & = P(C_{\epsilon} \in I(\mathcal{D}))  \geq 1-\delta, \label{eq:rate_problem_def}
\end{align}
where the left-hand side, denoted $\tilde{p}_{\epsilon}$, is referred to as the \textit{meta-probability} \cite{Angjelichinoski2019, kallehauge2022Globecom}. The one-sided confidence interval $I(\mathcal{D})$ thus provides the range of likely values for the $\epsilon$-outage capacity, and by choosing the minimum value in the interval, it is guaranteed with confidence $1-\delta$ that the outage probability is less than $\epsilon$. In addition to fulfilling \eqref{eq:rate_problem_def}, the selected rate should also maximize the throughput, which is measured by the \textit{normalized throughput}  \cite{kallehauge2022Globecom}
\begin{align}
    \tilde{R}_{\epsilon} =  \frac{R(\mathcal{D})(1-p_{\text{out}}(R(\mathcal{D}))}{C_{\epsilon}(1-\epsilon)}, \label{eq:throughput}
\end{align}
where the denominator is the throughput if the channel distribution were perfectly known, i.e., $\tilde{R}_{\epsilon}$ is exactly $1$ when $R(\mathcal{D}) = C_{\epsilon}$ \cite{Angjelichinoski2019}. 

The rate selection problem is hence solved using either of the methods described in Secs. \ref{sec:non_para}-\ref{sec:baseline} by selecting $R(\mathcal{D}) = I_{\min}(\mathcal{D})$ as the minimum value in the one-sided confidence interval. Note that the use of asymptotic results in all but the non-parametric baseline may cause the meta-probability $\tilde{p}_{\epsilon}$ to be lower than the target confidence $1-\delta$. 

\subsection{Numerical Results with Simulated Data: Urban Microcell} \label{subsec:quadriga}
\begin{figure}[t]
    \centering
        \includegraphics[scale = \picscale]{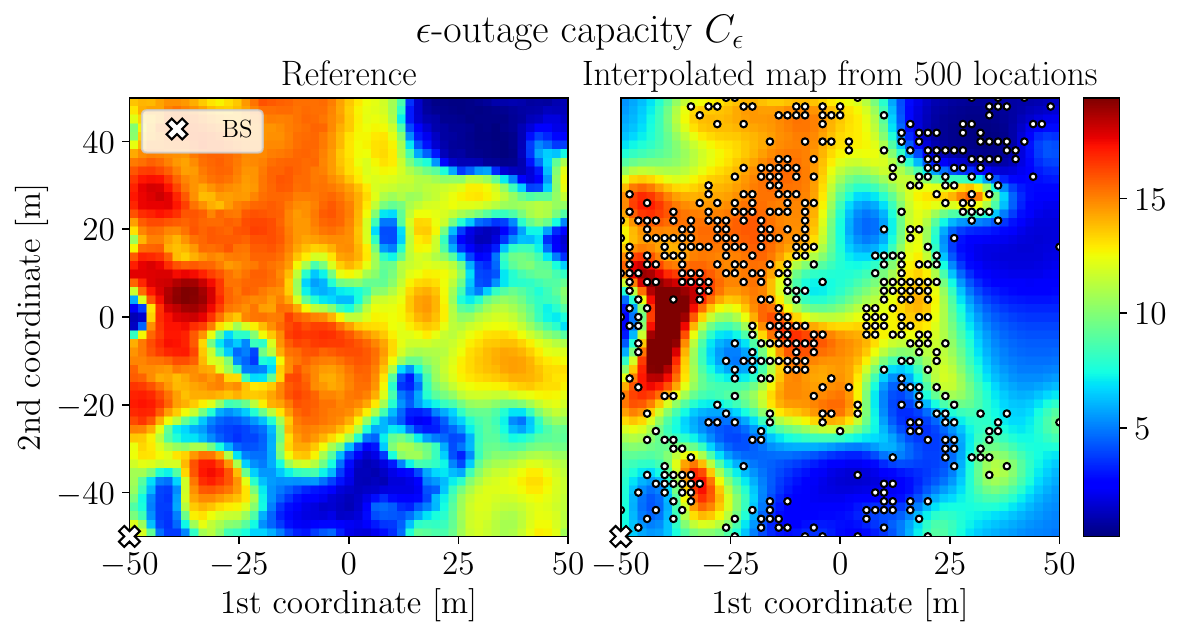}
        \includegraphics[scale = \picscale]{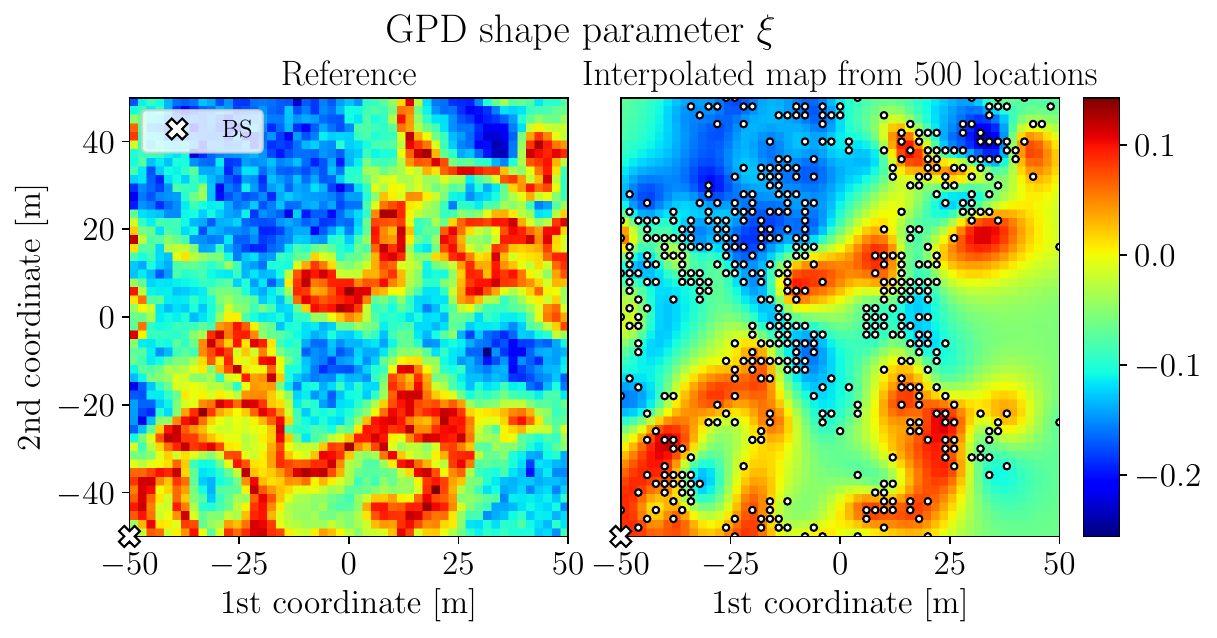}
    \caption{CDI maps of the $\epsilon$-outage capacity $C_{\epsilon}$ in bits/sec/Hz $\epsilon = 10^{-4}$ (top) and \gls{gpd} shape parameter $\xi$ (bottom) in a 3GPP Urban Microcell scenario. The left plots show the ground truth simulated data as a reference, and the right plots are the \gls{cdi}-maps interpolated based on $d = 500$ locations indicated by the dots. The \gls{cdi} maps are visualized as the \textit{mode} of the predictive distribution in \eqref{eq:pred_dist}, which for $C_{\epsilon}$ is $\exp(\mu(\mbf{s}) - \sigma^2(\mbf{s}))$ due to use of the log-transformation and and $\mu(\mbf{s})$ for $\xi$.}
        \label{fig:CDI_map_quadriga}
\end{figure}

The channel data in this section is simulated according to the 3GPP NR Urban Micro-Cell line-of-sight scenario using the simulation tool QuaDRiGa --- see \cite[p. 81]{quadriga} for further details. The simulated scenario has a cell area of $[-50,50] \times [-50,50]$ m$^2$ with user heights of $1.5$ m and the \gls{bs} at the corner of the cell at $\mbf{s}_{\text{BS}} = (-50,0,10)$ m. We simulate channels in a regular grid of points within the cell with grid size $\Delta s = 2$ m, yielding $2601$ points in total. The generated channel comprises $K = 20$ incoming paths, including the line-of-sight path, each one characterized by its magnitude and propagation delay. The central frequency is set to $3.6$ GHz, and the transmit power is fixed to $0$ dBm. Aiming at reproducing the impact of small-scale fading, an arbitrary number of channel realizations for each location in the grid are obtained by adding a random phase shift to each multipath component \cite{Molisch2002capacity}. Hence, the narrowband channel in \eqref{eq:rx_signal} is simulated stochastically as
\begin{align}
h =\sum_{k=1}^K a_k e^{j\theta_{k}}, \quad \theta_k \overset{\text{iid}}{\sim} \text{Unif}(-\pi,\pi) \label{eq:channel_quadriga}
\end{align}
where $a_k$ is the magnitude of the $k$-th multipath component simulated using Quadriga and $\theta_{k}$ is the random phase-change. With the drawn channel coefficients, samples of the instantaneous channel capacity are obtained as ${C = \log_2(1 + |h|^2/(BN_0))}$.
An estimate of the $\epsilon$-outage capacity $C_{\epsilon}$ with $\epsilon = 10^{-4}$ based on $10^6$ simulations for each grid point in the cell is depicted in Fig.~\ref{fig:CDI_map_quadriga}  (top left plot). 

The locations in dataset $\mathcal{D}$ in \eqref{eq:data} are generated as follows. Among the $2601$ grid points in the cell, the $d$ locations $\mbf{s}_1,\dots,\mbf{s}_d$ in $\mathcal{B}_{m,d}$ are drawn randomly without replacement along with an additional $d_{\text{test}}$ test locations. Each of the test locations represents an instance of $\mbf{s}_0$ where the rate should be selected. The $d + d_{\text{test}}$ locations are drawn jointly according to the method in \cite{kallehauge2022Globecom} of drawing from a clustered \textit{Thomas point process}, which mimics, for example, mobile users taking measurements along a busy street in an urban environment. The dots in Fig.~\ref{fig:CDI_map_quadriga} (right plots) illustrate the sampling process for the $d$ user locations used to form $\mathcal{B}_{m,d}$ (the $d_{\text{test}}$ test points are drawn similarly).
The figure also shows predicted \gls{cdi} maps of $C_{\epsilon}$ (top right) and the \gls{gpd} shape parameter $\xi$ (bottom right), which are used to obtain the prior for the two Bayesian approaches. It is seen that the Gaussian process is generally able to predict different channel statistics quite well except in locations with few observations, such as in the right part of the cell. Note, however, that the predicted \gls{cdi} map also models its uncertainty through the predictive variance, which is used directly in the rate selection methods to account for prediction errors. To increase the number of test points, the process of drawing $d + d_{\text{test}}$ locations is repeated $L$ times, thus generating a large amount of data to capture the spatial randomness in the scenario. We use $d = 500$, $d_{\text{test}} = 200$, and $L = 50$, giving a total of $d_{\text{test}}\cdot L = 10^{4} $ test locations for numerical results.

For each of the test locations, we select the communication rate $R$ using the two Bayesian methods and two baselines with target \gls{pep} $\epsilon = 10^{-4}$, confidence $1-\delta = 95\%$, $m = 10^6$ observations at each of the $d$ locations in $\mathcal{B}_{m,d}$, and a varying number of observations $n$ between $0$ and $10^6$ at the test locations. Following the heuristic approach in App. \ref{subsec:zeta_heuristic}, $\zeta = 2\cdot 10^{-3}$ is found to be a good choice for a global fraction for threshold selection rule in \eqref{eq:threshold_zeta}. 

The results are first evaluated by analyzing the achieved outage probabilities ${p_{\text{out}}(R) = P(C \leq R)}$ across all test locations. Aiming to get a tight estimation, $N_{\text{ref}} = 10^8$ independent simulations of $C$ are generated as reference to estimate the achieved outage probability\footnote{Using the analytic result in \eqref{eq:non_para_analytical}, it was found that the estimated meta-probability of the non-parametric baseline differs from the analytic result with only up to $0.36$ percentage points when $N_{\text{ref}} = 10^8$. The error increases significantly with fewer simulations, e.g., an error of up to $6.48$ percentage points when $N_{\text{ref}} = 10^6$.}.
All simulation settings are summarized in Tab. \ref{tab:settings_quadriga}.% 
\begin{table}[t]
    \centering
    \caption{Simulation settings: 3GPP Quadriga Urban Microcell} \label{tab:settings_quadriga}
    \resizebox{\linewidth}{!}{
    \begin{tabular}{ccc}
    \textbf{Symbol} & \textbf{Description}  & \textbf{Value} \\
    \hline & & \\[-1.5ex]
    $\mathcal{R}$ & Cell area & $[-50,50] \times [-50,50]$ m$^2$ \\
    $s_{\text{BS}}$ & \gls{bs} location & $(-50,0,10)$ m \\
    $\Delta s$ & Simulation grid size & $2$ m \\
    $BN_0$ & Noise level & $-90$ dBm \\
    $P_{\text{tx}}$ & Transmit power & $0$ dBm \\
    $f_c$ & Central frequency & $3.6$ GHz \\
    $K$ & \#Multipaths & 20 \\
    $d$ & \#Previous observed locations & $500$ \\
    $d_{\text{test}}$ & \#Test locations  & $200$ \\
    $L$ & \#Redrawings & $50$ \\ 
    $N_{\mathcal{D}}$ & \#Simulations used in $\mathcal{D}$ & $10^{6}$\\
    $N_{\text{ref}}$ & \#Simulations used to eval. results & $10^{8}$\\
    $\epsilon$ & Target \gls{pep} & $10^{-4}$ \\
    $1 - \delta$ & Target confidence & $95$ \% \\
    $\zeta$ & Threshold fraction in \eqref{eq:threshold_zeta} & $2\cdot 10^{-3}$ \\
    $r_{\min}$ & Minimum \#samples in \eqref{eq:threshold_zeta} & $100$ \\
    $T$ & \#Simulations in Alg. \ref{alg:metropolis_within_gibbs} & $10^4$ \\
    \end{tabular}}
\end{table}

\begin{figure}[t]
    \centering
    \includegraphics[scale = \picscale]{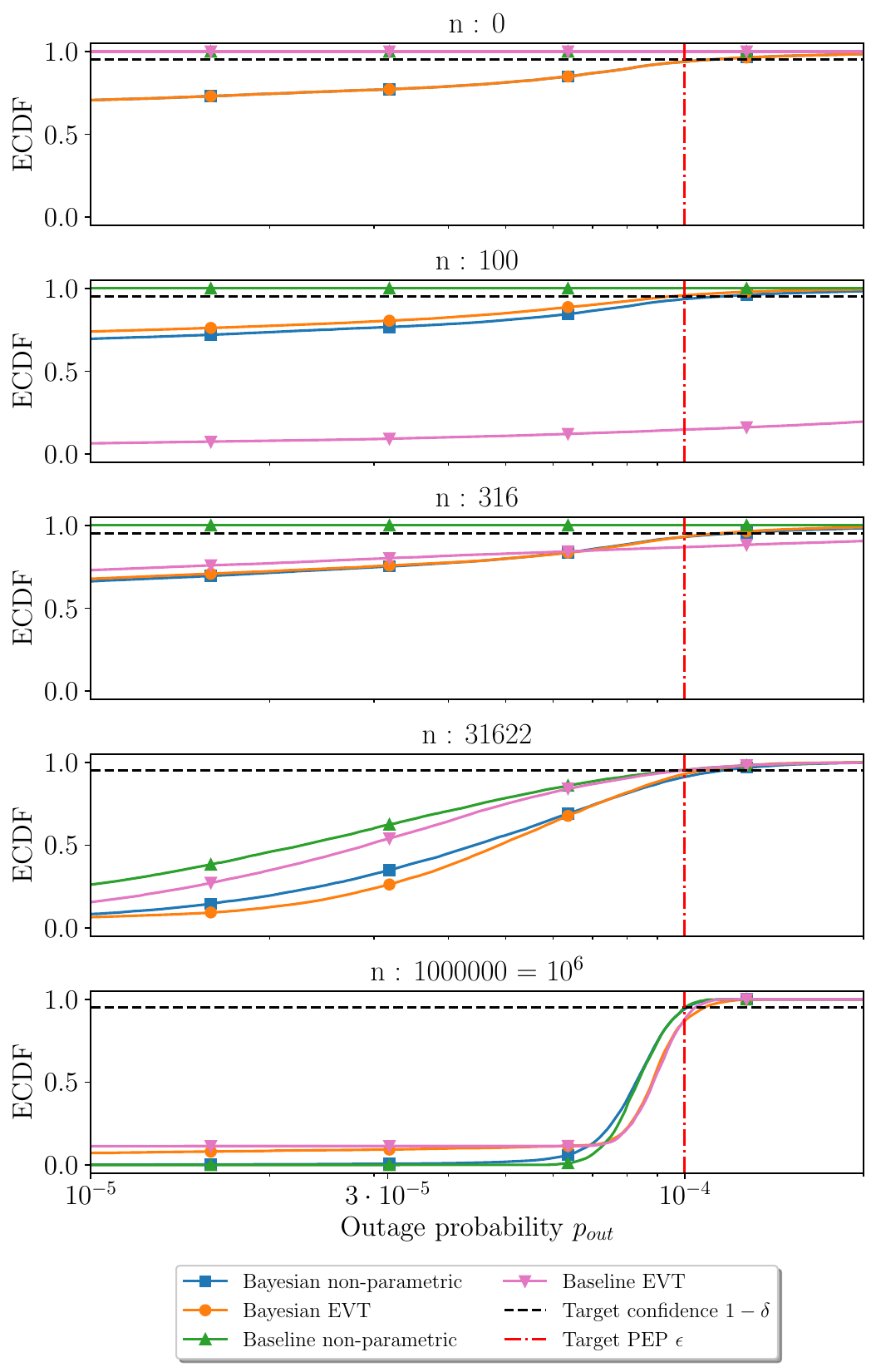}
    \caption{Outage probability in a 3GPP Urban Microcell scenario for the Bayesian non-parametric method, Bayesian EVT method, and the two baselines with varying number of observations per location $n$. The target \gls{pep} $\epsilon = 10^{-4}$ and confidence $1-\delta = 95\%$ are shown as dashed lines.}
    \label{fig:p_out_quadriga}
\end{figure}

\begin{figure}[t]
    \centering
    \includegraphics[scale = \picscale]{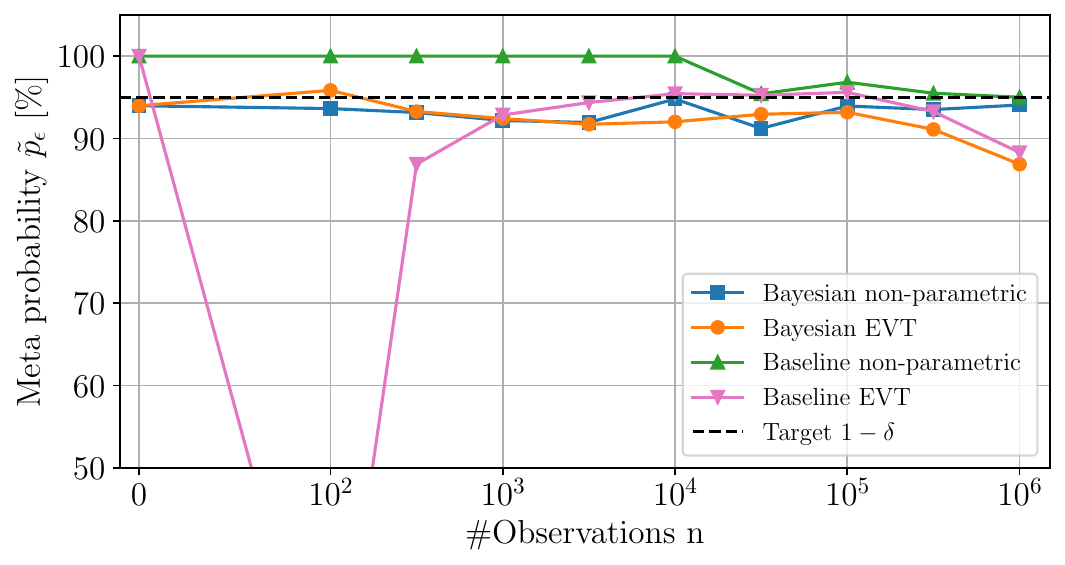}
    \caption{Meta probabilities in a 3GPP Urban Microcell scenario with varying number of observations per location $n$.}
    \label{fig:p_meta_quadriga}
\end{figure}

Fig. \ref{fig:p_out_quadriga} shows the \gls{ecdf} of the outage probability aggregated across locations with the number of observations $n$ between $0$ and $10^6$. 
It is seen that when no observations are available at the user location $\mbf{s}_0$, i.e., $n = 0$, the baselines have no information and select $R = 0$, as seen in the top plot where the outage probability is always $0$. The Bayesian methods, on the other hand, can rely on the \gls{cdi} maps despite no observations at the user location, although the rate is selected quite conservatively due to limited information, resulting in the outage probability being lower than required, i.e., $p_{\text{out}}(R) \ll \epsilon$ in most cases. The \gls{evt} baseline selects a non-zero rate given $n \geq 100$, although only $n = 100$ observations are seen to be insufficient, resulting in much higher outage probabilities than required. The Bayesian approaches, on the other hand, are still close to the target coverage for $n = 100$ due to their reliance on prior information. As $n$ increases, the outage probability of all methods gradually converges towards the target \gls{pep}\footnote{The non-parametric baseline requires $n \geq 29,956 \approx 3\cdot 10^4$ in order to select a non-zero rate with $\epsilon = 10^{-4}$ and $\delta = 5\%$ according to \eqref{eq:non_para_analytical}.}.
The value where the \glspl{ecdf} of the outage probability crosses $\epsilon = 10^{-4}$ corresponds to the meta-probability in \eqref{eq:rate_problem_def} as shown in Fig. \ref{fig:p_meta_quadriga}. 
It is seen that the meta-probabilities for all methods are approximately within  $\pm 5\%$ of the target coverage of $95\%$ except for the \gls{evt} baseline when $n = 100$. The non-parametric methods tend to be closer to the target, particulary for larger $n$, due to the fact that they are asymptotically exact.
Conversely, the methods based on \gls{evt} start to diverge from the target coverage as $n$ increases. This behavior is explained by the considerations in  Sec. \ref{sec:bias_discuss}: As the number of observations increases, any non-zero bias between the selected rate $R$  and outage capacity $C_{\epsilon}$ will cause the coverage to go towards $0$ or $100 \%$. The \gls{evt} methods will likely have a bias, and although the meta-probabilities are still close to $95\%$, we see evidence of this behavior for $n > 10^5$, where the meta-probability starts to trend towards $0\%$. 

\begin{figure}[t]
    \centering
    \includegraphics[scale = \picscale]{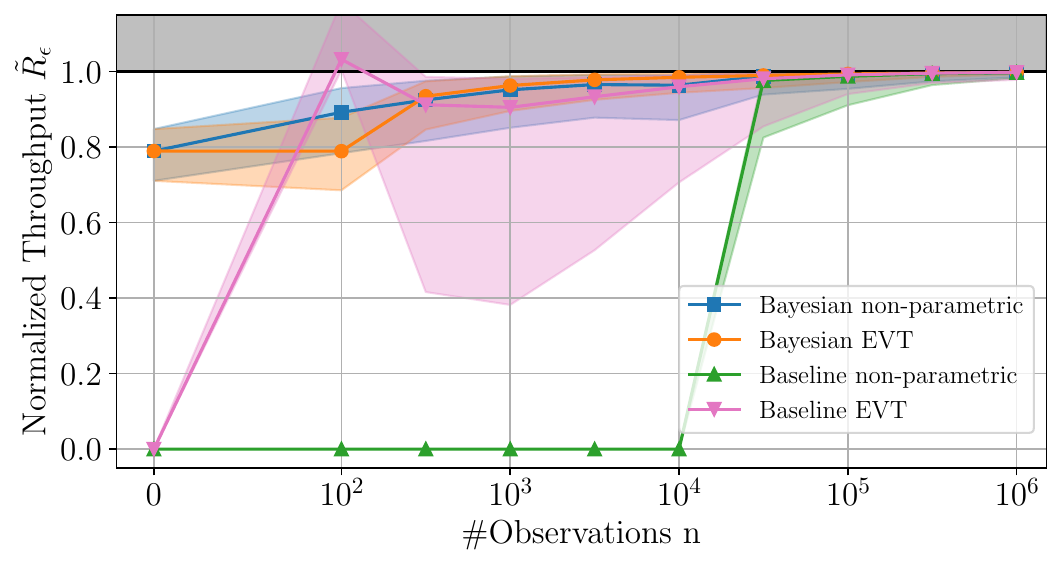}
    \caption{Normalized throughput in a 3GPP Urban Microcell scenario with varying number of observations per location $n$. The lines mark the median, while the shaded regions are between the 1st and 3rd quartile across all simulations.}
    \label{fig:throughput_quadriga}
\end{figure}

The normalized throughput in \eqref{eq:throughput} is evaluated and shown in Fig. \ref{fig:throughput_quadriga} as the 1st quartile (Q1), median (Q2), and 3rd quartile (Q3) of $\tilde{R}_{\epsilon}$ across the simulations. 
The gray region corresponds to normalized throughput greater than $1$ as seen for the \gls{evt} baseline with $n = 100$, which is generally undesirable since it also implies that the outage requirement is violated. For the remaining points, it is clear that the (normalized) throughput of the Bayesian methods is superior to the baselines with median values above $0.79$ for all $n$. With the exception of $n = 100$, the median throughput of the two Bayesian methods is generally comparable. The 1st quartile, however, is higher for the Bayesian \gls{evt} approach than the Bayesian non-approach, e.g., $0.94$ versus $0.87$ at $n = 10^4$. The \gls{evt} baseline has a median throughput comparable to the Bayesian solutions for $n \geq 316$, but a significantly lower 1st quartile starting from $0.4$ and slowing converging to $1$. The normalized throughput of the non-parametric baseline eventually converges to one, but it is generally the lowest and only takes non-zero values given an excessive number of observations, i.e., $n > 3\cdot 10^4$. 

\begin{figure}
     \centering
     \includegraphics[scale = \picscale]{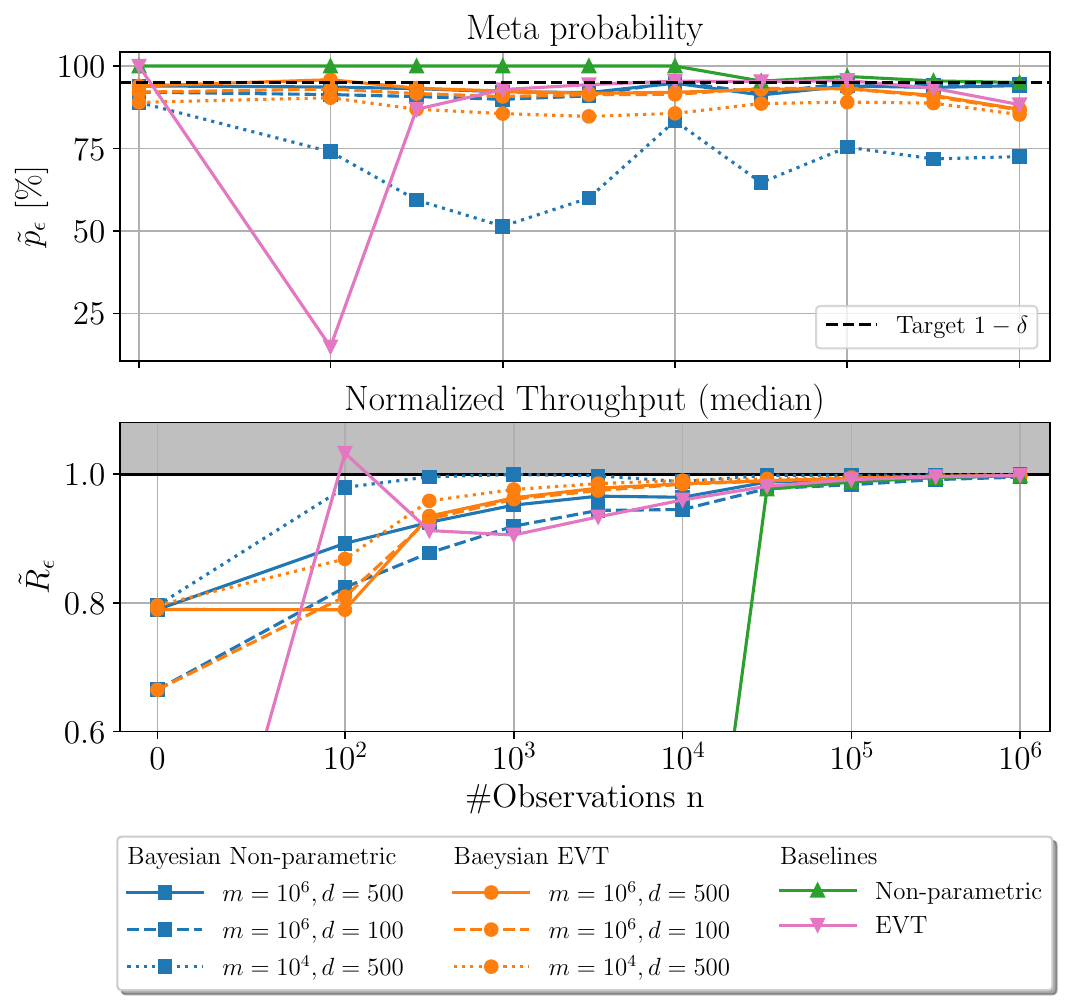}
     \caption{Meta-probability and median normalized throughput in a 3GPP Urban Microcell scenario with a different number of observations per location $n$, number of previous locations $d$, and number observations per previous location $m$.}
     \label{fig:quadriga_settings}
 \end{figure}

The effect of changing the number of previous observations in the cell in the dataset, i.e., $\mathcal{B}_{m,d}$, is also investigated. Fig. \ref{fig:quadriga_settings} shows the meta-probability and (median) normalized throughput when lowering the number of previously observed locations $d$ from $500$ to $100$ and when lowering the number of observations per previous location $m$ from $10^6$ to $10^4$. 
It is seen that lowering $d$ does not significantly affect the reliability of either of the Bayesian methods, although the throughput is generally lowered. On the other hand, lowering $m$ decreases the meta-probability of the non-parametric Bayesian approach with a minimum of $51.4\%$ in the simulations, whereas the \gls{evt} Bayesian approach is affected less with a minimum of $84.7\%$. Hence, despite reducing the number of previous observations, the Bayesian approaches still have superior throughput and comparable reliability with respect to the baselines (for the \gls{evt} approach), showing that even a smaller collection of previous measurements can be used advantageously. 

\subsection{Numerical Results with Experimental Data: Rich Scattering Environment} \label{subsec:apms}
\begin{figure}[t]
    \centering
        \includegraphics[scale = \picscale]{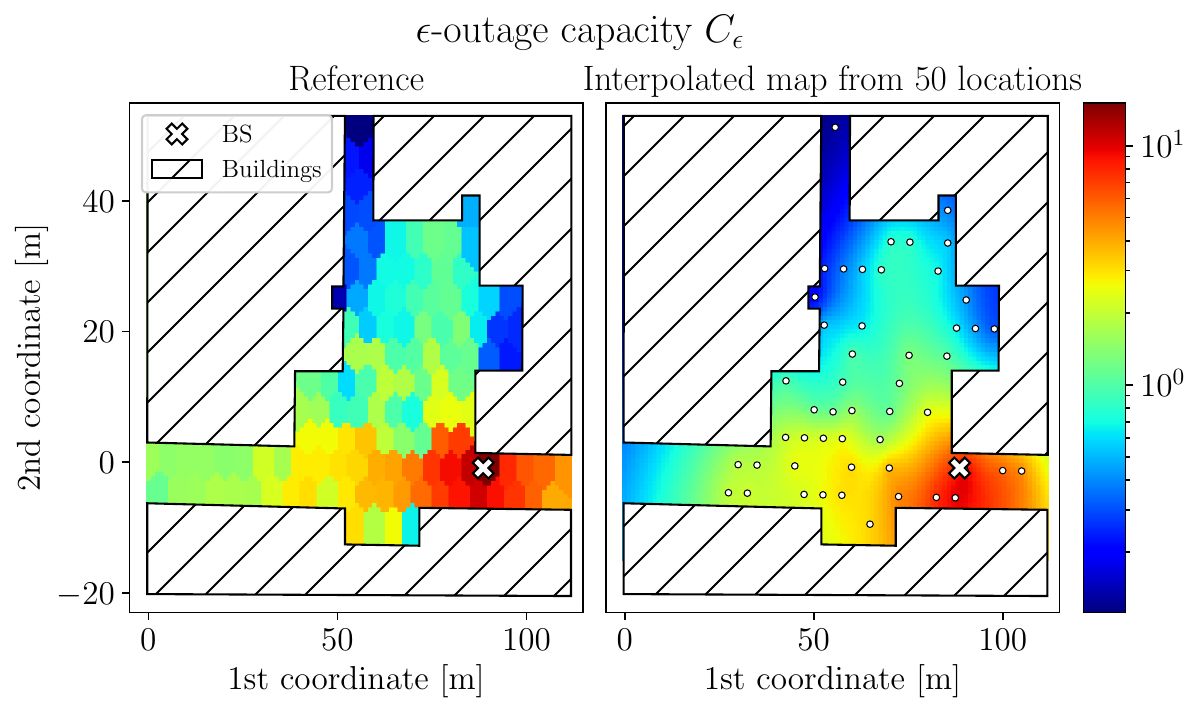}
        \includegraphics[scale = \picscale]{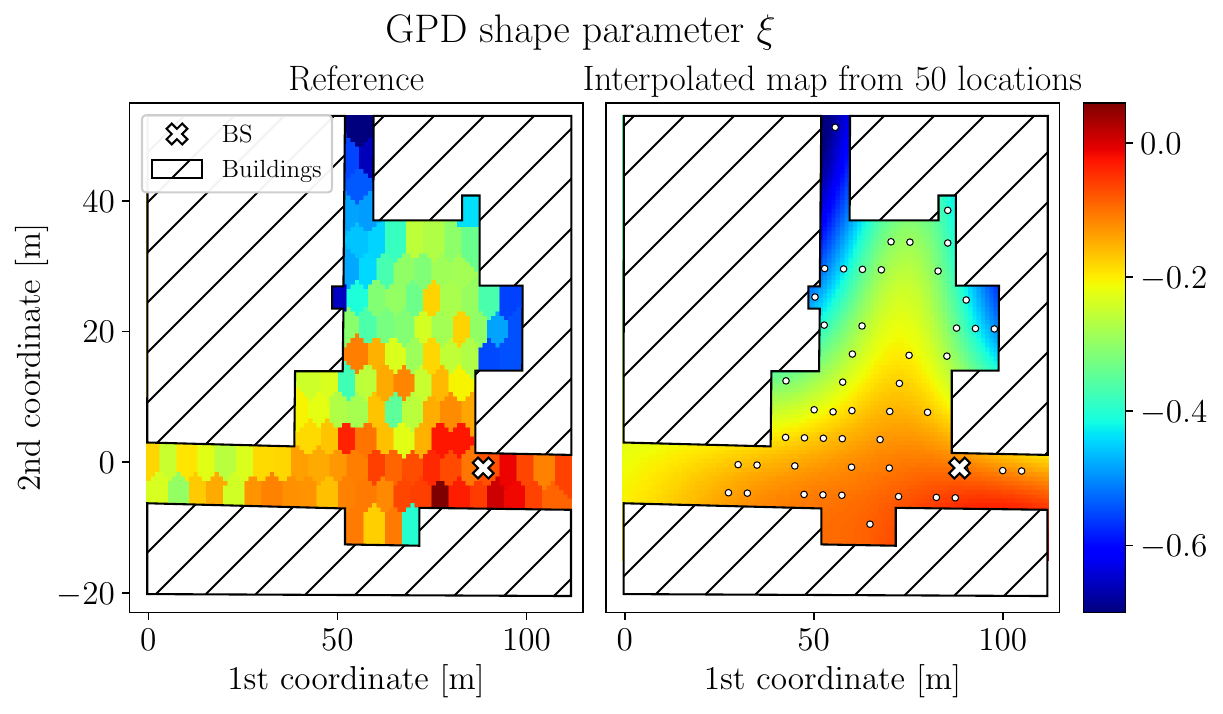}
    \caption{Similar to Fig. \ref{fig:CDI_map_quadriga} with CDI maps of the $\epsilon$-outtage capacity $C_{\epsilon}$ in bits/sec/Hz with $\epsilon = 10^{-2}$ (top) and \gls{gpd} shape parameter $\xi$ (bottom) in an experimentally measured rich scattering environment. The \gls{cdi}-maps on the right are based on $d = 50$ locations indicated by the dots.}
    \label{fig:CDI_map_APMS}
\end{figure}

The data in this section consists of channel-sounding measurements in a rich scattering environment with a fixed receiver (i.e., \gls{bs}) and a transmitter (i.e., user) positioned at $127$ different locations in a triangular grid with sidelength $\Delta s = 5$ m --- we refer the reader to \cite{kallehauge2024experimental} for additional details on the measuring campaign and used equipment. The channel-sounding measurements contain $8001$ channel estimates from $2$ to $10$ GHz, hence, a $1$ MHz difference between estimates. Each of these samples can be interpreted as a realization of random fading, either due to random frequency hopping or, as argued in \cite{kallehauge2024experimental}, an effect \textit{similar} to small-scale fading induced by small movements of the transmitter or receiver. The dataset hence provides $8001$ measurements for each of $127$ locations, which are treated as independent fading measurements of the channel $h$ in \eqref{eq:rx_signal}. The instantaneous capacity is obtained as $C = \log_2(1 + |h|^2/(BN_0))$ from the measurements assuming $BN_0 = -70$ dBm. Fig. \ref{fig:CDI_map_APMS} (top left) illustrates the estimated $\epsilon$-quantile of the fading measurements for $\epsilon = 10^{-2}$. It is seen that the channel has a somewhat smooth spatial dependency but with some sudden drops in signal level, e.g., close to corners of buildings.  

The rate-selection methods are evaluated similarly to the previous section, with a few exceptions due to the limited size of the dataset. Firstly, rather than drawing points from a clustered point process, we simply draw $d = 50$ locations uniform randomly without replacement among the $127$ available as the previous user locations in $\mathcal{D}$ and then use the remaining $d_{\text{test}} = 77$ points as test locations. Fig. \ref{fig:CDI_map_APMS} (right plots) illustrate the sampling process and shows predicted \gls{cdi} maps of $C_{\epsilon}$ (top right) and the \gls{gpd} shape parameter $\xi$ (bottom right). Again, it is seen that the Gaussian process is generally able to predict different channel statistics quite well as long as they vary smoothly in space. The process of drawing points is repeated $L = 130$ times, giving a total of $d_{\text{test}}\cdot L = 10010$ test locations for results. For each of the re-draws, the $8001$ measurements are divided randomly into $N_{\mathcal{D}} = 4000$ observations used to construct the dataset  $\mathcal{D}$ and $N_{\text{ref}} = 4001$ reference observations used to evaluate the results. We select a target \gls{pep} of $\epsilon = 10^{-2}$ with confidence $1 - \delta = 95\%$. The dataset $\mathcal{D}$ has $m = 4000$ observations at each of the previous locations and a varying number of observations  $n$ between $0$ and $4000$ at the test locations. The fraction $\zeta$ in \eqref{eq:threshold_zeta} is chosen to $0.4$ based on the heuristic in App. \ref{subsec:zeta_heuristic}. All simulation settings are summarized in Table \ref{tab:settings_apms}.
\begin{table}[t]
    \centering
    \caption{Simulation settings: Experimental Rich Scattering} \label{tab:settings_apms}
    %\resizebox{\linewidth}{!}{
    \begin{tabular}{ccc}
    \textbf{Symbol} & \textbf{Description}  & \textbf{Value} \\
    \hline & & \\[-1.5ex]
    - & Max. dist. from rx to tx & $ 88.6$ m \\
    $\Delta s$ & Triangular grid size & $5$ m \\
    $BN_0$ & Noise level & $-70$ dBm \\
    $P_{\text{tx}}$ & Transmit power & $0$ dBm \\
    $d$ & \#Previous observed locations & $50$ \\
    $d_{\text{test}}$ & \#Test locations  & $77$ \\
    $L$ & \#Redrawings & $130$ \\ 
    $N_{\mathcal{D}}$ & \#Simulations used in $\mathcal{D}$ & $4000$\\
    $N_{\text{ref}}$ & \#Simulations used to eval. results & $4001$\\
    $\epsilon$ & Target \gls{pep} & $10^{-2}$ \\
    $1 - \delta$ & Target confidence & $95$ \% \\
    $\zeta$ & Threshold fraction in \eqref{eq:threshold_zeta} & $0.4$ \\
    $r_{\min}$ & Minimum \#samples in \eqref{eq:threshold_zeta} & $50$ \\
    $T$ & \#Monte Carlo samples in Alg. \ref{alg:metropolis_within_gibbs} & $10^4$ \\
    \end{tabular}%}
\end{table}

\begin{figure}[t]
    \centering
    \includegraphics[scale = \picscale]{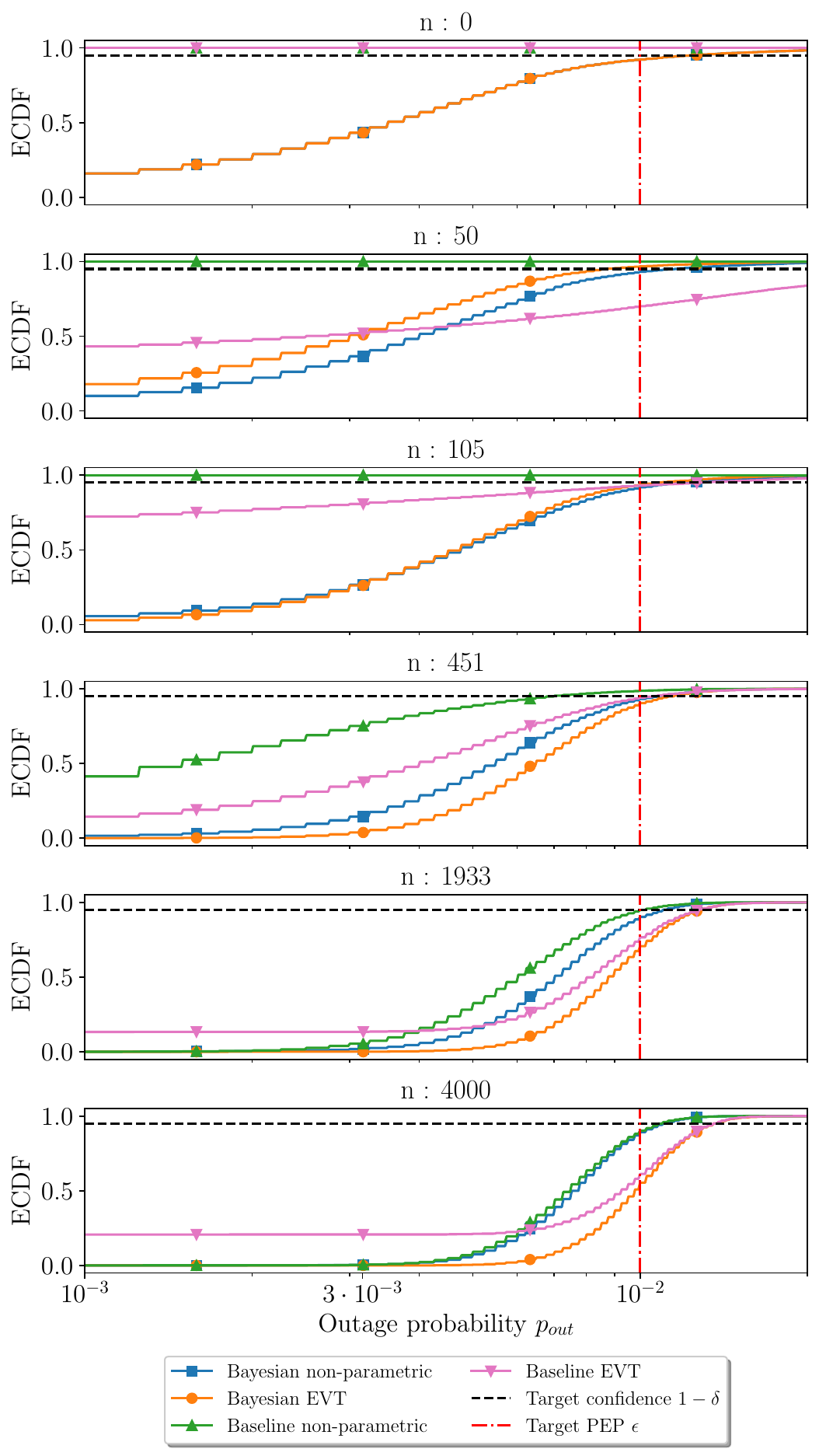}
    \caption{Outage probability with experimental data for the Bayesian non-parametric method, Bayesian EVT method, and the two baselines with varying number of observations per location $n$. The target \gls{pep} $\epsilon = 10^{-2}$ and confidence $1-\delta = 95\%$ as shown as dashed lines.}
    \label{fig:p_out_apms}
\end{figure}

\begin{figure}[t]
    \centering
    \includegraphics[scale = \picscale]{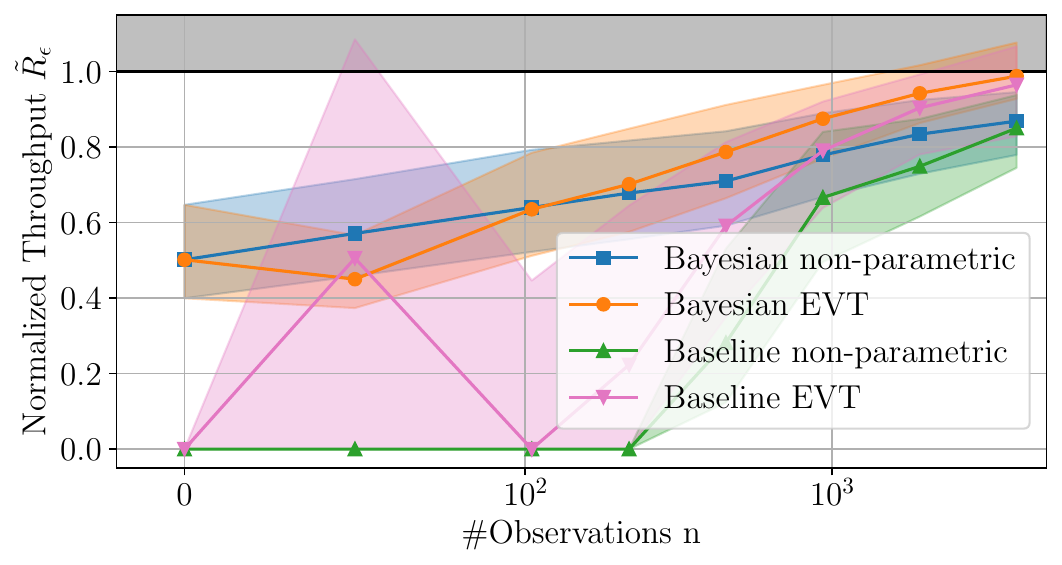}
    \caption{Normalized throughput with experimental data with varying number of observations per location $n$. The lines mark the median, while the shaded regions are between the 1st and 3rd quartile across all simulations.}
    \label{fig:throughput_apms}
\end{figure}

Fig. \ref{fig:p_out_apms} shows the \gls{ecdf} of the outage probability aggegated across locations. 
The results for the experimental data are overall similar to simulated channels in the previous section, with the Bayesian approaches generally being closer to the target \gls{pep} when $n$ is low, and all methods gradually converge to the target as $n$ increases. One interesting observation is in the high-observation domain, where the \gls{ecdf}-probabilities of the \gls{evt} approaches at the target $\epsilon = 10^{-2}$ start to deviate from the target confidence of $95\%$, as opposed to the non-parametric approaches. This effect is again explained by Sec. \ref{sec:bias_discuss}, where even a small model bias (here from using \gls{evt}) will cause the coverage guarantee to converge to either $0$ or $100$ \% as $n$ increases. However, in the scenario with fewer observations, the outage probabilities are below the target $\epsilon$ with high confidence close to $95\%$. The only exception is the baseline \gls{evt}-approach with $n = 50$, similar to what we observed for the simulated channels in Fig. \ref{fig:p_out_quadriga}. With the limited size of the reference dataset, i.e., $N_{\text{ref}} = 4001$, accurate estimation of the meta-probability in \eqref{eq:rate_problem_def} is not possible due to the difficulty in evaluating the binary event $p_{\text{out}}(R) \leq \epsilon$ whenever $p_{\text{out}}(R) \approx \epsilon$ \footnote{For example, the meta-probability estimated from the reference data deviated with up to $5.47$ percentage points from the analytic result in \eqref{eq:non_para_analytical} for the non-parametric baseline.}. We, therefore, omit the estimated meta-probabilities and refer to Fig. \ref{fig:p_out_apms}, which indicate the approximate values at $p_{\text{out}} = 10^{-2}$. The normalized throughput is shown in Fig. \ref{fig:throughput_apms} as the 1st quartile, median, and 3rd quartile of $\tilde{R}_{\epsilon}$ across the simulations. 
Again, the results are similar to those from the simulated channels, with the Bayesian approaches having significantly higher throughput than the baselines, particularly in the low-sample domain.  

Some essential findings are made from the results with simulated and experimental data. First, the Bayesian approaches that rely on previous observations in the cell perform significantly better in terms of throughput while achieving comparable reliability guarantees in the important scenario when the number of observations of the new user location is low, i.e., either none or just a few hundred observations. Notice also that there are gains in terms of throughput even for very few observations, e.g., for the non-parametric approach with $n  = 100$ and $n = 50$ in the simulated and experimental data, despite modeling a rare outage event that happens only with probability $10^{-4}$ and $10^{-2}$, respectively. The normalized throughput eventually converges to $1$ for all methods as $n$ increases, although the non-parametric approaches are generally more accurate in achieving the desired reliability guarantee for high $n$. This clearly highlights the trade-off between reliability and throughput, where the non-parametric methods generally offer tighter reliability guarantees but at the cost efficiency compared to the \gls{evt} methods. One interesting solution, which balances these metrics, would be to use the Bayesian \gls{evt} approach in the low observation domain, e.g., when $n < 1/\epsilon$, and then rely on the Bayesian non-parametric approach in the high-observation domain.

\section{Conclusions} \label{sec:conclusion}
This paper studied how to leverage previous measurements from other locations in a wireless network in combination with new measurements to estimate rare-event statistics of a communication channel, which is critical in areas such as \gls{urllc}. The general problem was formulated based on statistical learning principles and asked how to infer a confidence interval for rare-event quantiles of the channel. This was later exemplified by the specific problem of location-aided rate selection in ultra-reliable communication under small-scale fading. 

Two novel Bayesian approaches using spatial prediction based on \gls{cdi} maps approaches were proposed, one using non-parametric statistics and one using \gls{evt}. These were compared to existing baselines in the rate-selection example, first with simulated channels from the QuaDRiGa simulator and then with experimentally measured channels. It was found that the Bayesian approaches for rate selection achieved significantly higher throughput than the baselines at comparable levels of reliability. When no channel measurements are available, the Bayesian approach can rely on spatial prediction to achieve non-zero throughput, as opposed to the baselines. The throughput gradually increases for all methods, but the Bayesian methods converge faster, particularly the approach using \gls{evt}. The non-parametric approaches have the advantage of tighter coverage guarantees in the high observation domain due to being unbiased, but using \gls{evt} is otherwise an improvement. It is thus concluded that previous channel measurement in a communication scenario can indeed be leveraged advantageously and that the proposed Bayesian estimation approaches can significantly increase the accuracy and efficiency of estimating rare-event statistics. Topics for future works include improving coverage guarantees, i.e., the probability that the rare-event statistics are included in the estimated confidence intervals, dealing with non-stationary propagation environments and dynamic updating of \gls{cdi} maps, improving threshold selection strategies for the \gls{evt}-based solutions, detection of blockages based on non-smoothness, and further experimental work to verify and improve upon the estimation methods for rare-event statistics.  
\appendices 
\section{Heuristics for Selecting Threshold Fraction $\zeta$} \label{subsec:zeta_heuristic}
The parameter $\zeta$ is used in \eqref{eq:threshold_zeta} as the fraction of measurements used to model the tail distribution. Simple heuristics such as selecting $\zeta = 10\%$ are often used \cite{scarrott2012review}, but may suffer due to not depending on the actual data. We suggest the following approach as a more advanced heuristic, which also incorporates information about the data. 

Given $n$ independent observations $\mbf{X}^n$, the aforementioned mean deficit plot can be used to infer the threshold $u$ from Thm. \ref{thm:pickands}. In particular, we have that if Thm. \ref{thm:pickands} applies for threshold $u_0$ with parameters $(\sigma_{u_0},\xi)$, then \cite{Mehrnia2022, coles2001introduction}
\begin{align}
    e(u) = E[u - X \cond X < u] = \frac{\sigma_{u_0} - \xi(u - u_0)}{1 - \xi} \label{eq:mean_deficit}
\end{align}
for all $u \leq u_0$, which is seen to be a linear function of the threshold $u$. Plotting numeric estimates of the mean deficit against different thresholds can reveal the interval of values $u$ where $e(u)$ behaves linearly, and the threshold can then be chosen as the largest value of this interval --- see \cite[p. 78-80]{coles2001introduction} for further details. Detecting the threshold from the estimated mean deficit is usually done manually. Given a threshold $u$ selected using the mean deficit plot, the corresponding fraction can be computed as $\zeta = n^{-1} \sum_{i=1}^n \mbf{1}(X_i \leq u)$. This process is repeated for channels measured at $d_{\text{cal}}$ different locations giving a set of fractions $\zeta_1,\dots,\zeta_{d_{\text{cal}}}$. To balance the bias-variance trade-off, we propose the heuristic of selecting the \textit{median} of the computed fractions and using this as a global fraction. This calibration process only needs to be done once and is, hence, more practical with the critical assumption that the dataset used for calibration is representative of the entire cell.

\ifCLASSOPTIONcaptionsoff
  \newpage
\fi

\bibliographystyle{ieeetr}
\bibliography{references}

\begin{thebibliography}{10}

\bibitem{kalor24massivecritical}
A.~E. Kal{\o}r {\em et~al.}, ``Wireless {6G} connectivity for massive number of
  devices and critical services,'' {\em arXiv:2401.01127}, 2024.

\bibitem{Angjelichinoski2019}
M.~Angjelichinoski, K.~F. Trillingsgaard, and P.~Popovski, ``A statistical
  learning approach to ultra-reliable low latency communication,'' {\em IEEE
  Trans. Commun.}, vol.~67, no.~7, pp.~5153--5166, 2019.

\bibitem{kallehauge2022Globecom}
T.~Kallehauge, P.~Ramírez-Espinosa, A.~E. Kalør, C.~Biscio, and P.~Popovski,
  ``Predictive rate selection for ultra-reliable communication using
  statistical radio maps,'' in {\em IEEE Conf. Global Telecommun. Conf.
  (GLOBECOM)}, pp.~4989--4994, 2022.

\bibitem{kallehauge2024experimental}
T.~Kallehauge, A.~E. Kal{\o}r, F.~Zhang, and P.~Popovski, ``Experimental study
  of spatial statistics for ultra-reliable communications,'' {\em
  arXiv:2402.11356}, 2024.

\bibitem{Mehrnia2022}
N.~Mehrnia and S.~Coleri, ``Wireless channel modeling based on extreme value
  theory for ultra-reliable communications,'' {\em IEEE Trans. Wirel. Commun.},
  vol.~21, no.~2, pp.~1064--1076, 2022.

\bibitem{gomes2022rare}
A.~Gomes, J.~Kibiłda, and L.~A. DaSilva, ``Capturing rare network conditions
  to dimension resources for ultra-reliable communication,'' {\em IEEE Commun.
  Lett.}, vol.~26, no.~11, pp.~2789--2793, 2022.

\bibitem{perez2023extreme}
D.~E. Pérez, O.~L.~A. López, and H.~Alves, ``Extreme value theory-based
  robust minimum-power precoding for {URLLC},'' {\em IEEE Trans. Wirel.
  Commun., early access}, May 2024.

\bibitem{Mehrnia2022confidence}
N.~Mehrnia and S.~Coleri, ``Incorporation of confidence interval into rate
  selection based on the extreme value theory for ultra-reliable
  communications,'' in {\em Eur. Conf. Netw. Commun. {6G} Sum. (EuCNC/6G
  Summit)}, pp.~118--123, 2022.

\bibitem{Chowdappa2018}
V.-P. Chowdappa, C.~Botella, J.~J. Samper-Zapater, and R.~J. Martinez,
  ``Distributed radio map reconstruction for {5G} automotive,'' {\em {IEEE}
  Intell. Transp. Syst. Mag.}, vol.~10, no.~2, pp.~36--49, 2018.

\bibitem{Lima2021Sensing}
C.~De~Lima {\em et~al.}, ``Convergent communication, sensing and localization
  in {6G} systems: {An} overview of technologies, opportunities and
  challenges,'' {\em IEEE Access}, vol.~9, pp.~26902--26925, 2021.

\bibitem{Studer2018}
C.~Studer, S.~Medjkouh, E.~Gonultaş, T.~Goldstein, and O.~Tirkkonen, ``Channel
  charting: Locating users within the radio environment using channel state
  information,'' {\em IEEE Access}, vol.~6, pp.~47682--47698, 2018.

\bibitem{kallehauge2023magazine}
T.~Kallehauge, A.~E. Kalør, P.~Ramírez-Espinosa, M.~Guillaud, and
  P.~Popovski, ``Delivering ultra-reliable low-latency communications via
  statistical radio maps,'' {\em IEEE Wirel. Commun.}, vol.~30, no.~2,
  pp.~14--20, 2023.

\bibitem{gonzalez2024integrated}
N.~Gonz{\'a}lez-Prelcic {\em et~al.}, ``The integrated sensing and
  communication revolution for {6G}: Vision, techniques, and applications,''
  {\em arXiv:2405.01816}, 2024.

\bibitem{zeng2024tutorial}
Y.~Zeng {\em et~al.}, ``A tutorial on environment-aware communications via
  channel knowledge map for {6G},'' {\em IEEE Commun. Surv. Tut., early
  access}, Feb. 2024.

\bibitem{Kulzer2021}
D.~F. Külzer, S.~Stańczak, and M.~Botsov, ``{CDI} maps: Dynamic estimation of
  the radio environment for predictive resource allocation,'' in {\em 2021 IEEE
  32nd Annu. Int. Symp Pers. Indor. Mob. Rad. Commun. (PIMRC)}, pp.~892--898,
  2021.

\bibitem{yu2020channel}
J.~Yu {\em et~al.}, ``Channel measurement and modeling of the small-scale
  fading characteristics for urban inland river environment,'' {\em IEEE Trans.
  Wirel. Commun.}, vol.~19, no.~5, pp.~3376--3389, 2020.

\bibitem{perez2024evt}
D.~E. P{\'e}rez, O.~L.~A. L{\'o}pez, and H.~Alves, ``{EVT}-enriched radio maps
  for {URLLC},'' {\em arXiv:2404.04558}, 2024.

\bibitem{kallehauge2023statistical}
T.~Kallehauge, M.~V. Vejling, P.~Ramírez-Espinosa, K.~Kansanen, H.~Wymeersch,
  and P.~Popovski, ``On the statistical relation of ultra-reliable wireless and
  location estimation,'' {\em IEEE Trans. Wirel. Commun., early access}, Feb.
  2024.

\bibitem{Ord1994}
A.~Stuart and O.~J. Keith, {\em Kendall's Advanced Theory of Statistics:
  Distribution Theory}, vol.~1 of {\em Kendall's Library of Statistics}.
\newblock Halsted Press and Wiley \& Sons, New York, 6~ed., 1994.

\bibitem{Kay1998}
S.~M. Kay, {\em Fundementals of Statistical Signal Processing}, vol.~II.
\newblock Pearson, 1998.

\bibitem{coles2001introduction}
S.~Coles, J.~Bawa, L.~Trenner, and P.~Dorazio, {\em An introduction to
  statistical modeling of extreme values}, vol.~208.
\newblock Springer, 2001.

\bibitem{balkema1974residual}
A.~A. Balkema and L.~De~Haan, ``Residual life time at great age,'' {\em Annu.
  Prob.}, vol.~2, no.~5, pp.~792--804, 1974.

\bibitem{scarrott2012review}
C.~Scarrott and A.~{MacDonald}, ``A review of extreme value threshold
  estimation and uncertainty quantification,'' {\em {REVSTAT}-Statist. J.},
  vol.~10, no.~1, pp.~33--60, 2012.

\bibitem{bishop2006pattern}
C.~M. Bishop and N.~M. Nasrabadi, {\em Pattern recognition and machine
  learning}, vol.~4.
\newblock Springer, 2006.

\bibitem{yang14quasistatic}
W.~Yang, G.~Durisi, T.~Koch, and Y.~Polyanskiy, ``Quasi-static multiple-antenna
  fading channels at finite blocklength,'' {\em IEEE Trans Inf. Theor.},
  vol.~60, no.~7, pp.~4232--4265, 2014.

\bibitem{quadriga}
Fraunhofer Heinrich Hertz Institute, {\em Quasi Deterministic Radio Channel
  Generator User Manual and Documentation}, v2.4.0~ed., October 2020.
\newblock Accessed: 19. Jun. 2024. [Online]. Available:
  \href{https://quadriga-channel-model.de/}{https://quadriga-channel-model.de/}.

\bibitem{Molisch2002capacity}
A.~Molisch, M.~Steinbauer, M.~Toeltsch, E.~Bonek, and R.~Thoma, ``Capacity of
  {MIMO} systems based on measured wireless channels,'' {\em IEEE J. Sel. Areas
  Commun. (JSAC)}, vol.~20, no.~3, pp.~561--569, 2002.

\end{thebibliography}

\end{document}